\documentclass[12pt,onecolumn]{article}
\usepackage{latexsym}
\usepackage{amsmath,url}
\usepackage[colorlinks=true,citecolor=black,linkcolor=black]{hyperref}
\usepackage{amstext,amsfonts,epsf,epsfig,pifont}
\usepackage{graphicx,bm,amsmath}
\usepackage[all=normal,floats,leading,lists,sections]{savetrees}
\usepackage[hmargin=.8in,vmargin=0.8in]{geometry}

\usepackage{url}

\usepackage{mathtools} 
\usepackage{latexsym}
\usepackage{amsmath}
\usepackage{amstext,amsfonts,epsf,epsfig,pifont}
\usepackage{graphics,graphicx,bm,amsmath,multicol}

\newlength{\intwidth}

\newcommand{\rd}{\ensuremath{\mathrm{d}}}
\newcommand{\ri}{\ensuremath{\mathrm{i}}}
\newcommand{\re}{\ensuremath{\mathrm{e}}}

\newcommand{\E}{\ensuremath{\mathrm{E}}}

\newcommand{\bx}{\ensuremath{\mathbf{x}}}
\newcommand{\by}{\ensuremath{\mathbf{y}}}

\newcommand{\calE}{\ensuremath{\mathcal{E}}}

\usepackage[comma,square,numbers]{natbib}

\begin{document}


\title{\footnotesize The definitive version of this article has been published by the Royal Society of London. \\
Lilly, J. M. (2017). Element analysis: a wavelet-based method for analysing time-localized events in noisy time series.  \emph{Proc. R. Soc. A}, 20160776, \url{http://dx.doi.org/10.1098/rspa.2016.0776}. \\\vspace{.05in} \Large Element analysis: a wavelet-based method \\ for analyzing time-localized events in noisy time series}%
\author{J. M. Lilly\footnote{NorthWest Research Associates, Redmond, Washington, USA.}}

\maketitle

\begin{abstract}
A method is derived for the quantitative analysis of signals that are composed of superpositions of isolated, time-localized ``events''.  Here these events are taken to be well represented as rescaled and phase-rotated versions of generalized Morse wavelets, a broad family of continuous analytic functions.   Analyzing a signal composed of replicates of such a function using another Morse wavelet allows one to directly estimate the properties of events from the values of the wavelet transform at its own maxima.  The distribution of events in general power-law noise is determined in order to establish significance based on an expected false detection rate.  Finally, an expression for an event's ``region of influence'' within the wavelet transform permits the formation of a criterion for rejecting spurious maxima due to numerical artifacts or other unsuitable events.  Signals can then be reconstructed based on a small number of isolated points on the time/scale plane. This method, termed {\em element analysis}, is applied  to the identification of long-lived eddy structures in ocean currents as observed by along-track  measurements of sea surface elevation from satellite altimetry. 
\end{abstract}

\section{Introduction}

A common problem in time series analysis is the need to detect and describe signals that are non-sinusoidal in nature.   In such cases, continuous wavelet analysis provides an attractive alternative to Fourier analysis.  For signals that are close to  being sinusoidal, a method known as ``wavelet ridge analysis'' \cite{delprat92-itit,mallat,lilly06-npg,lilly09-asilomar,lilly10-itit,lilly12a-itsp}  provides a powerful tool for  detection and quantitative analysis.  At the other extreme, for signals that are nearly singular in nature, the ``modulus maxima'' method \cite{mallat92-itit,mallat}, has proved useful.  These popular methods represent the signal as being  supported entirely on nearly horizontal, or nearly vertical, curves on the time/scale plane, respectively.

A third class of signals is neither nearly sinusoidal nor nearly singular, but is composed of self-similar events that are localized in time and that may be considered as barely oscillatory or even non-oscillatory.  That is, the signal is considered to be composed of isolated events that themselves resemble wavelets.  In contrast to the wavelet ridges and the modulus maxima curves, signals of this type are supported only at isolated points distributed, like stars or dust, sparsely throughout the time/scale plane.  Because individual wavelets are good approximations for phenomena ranging from heartbeats recorded by an electrocardiogram to propagating wave packets to climate oscillations, one may expect  signals of this type to be fairly widespread.

A particular example comes from oceanography, and involves satellite observations of so-called ``coherent eddies'', swirling $\mathcal{O}$(10--100)~km vortex structures that are ubiquitous features of the ocean circulation.  Such features, which are frequently modeled as having sea surface height anomalies that are Gaussian in shape, are observed along the narrow ground tracks of satellite altimeter instruments.  This leads to time series in which nearly Gaussian bumps or depressions of varying scales are embedded, together with noise as well as other sources of sea surface height variability.   While such ``along-track'' observations are occasionally used to study eddies \cite{byrne95-jpo,zavala-hidalgo03-jpo,lilly03-pio,lehenaff14-pio}, a far more common approach, as in the watershed study of \cite{chelton11-pio}, is to rely on mapped data products.  Because the altimeter records typically have about 5~km resolution in the along-track direction, but about 100~km resolution in the cross-track direction, the creation of mapped fields involves a horizontal smoothing that reduces the along-track resolution by an order of magnitude.  
 
Inspired by this problem, yet imagining that its solution may be of general interest, the following model for a time series is proposed.  The real-valued time series $x(t)$ is represented as containing time-offset, phase-shifted, and rescaled copies of some time-localized complex-valued function $\psi(t)$, together with measurement noise that is assumed to be Gaussian and stationary,
\begin{equation}\label{signalmodel}
x(t) = \sum_{n=1}^N \Re\left\{c_n \psi\left(\frac{t-t_n}{\rho_n}\right)\right\} +x_\epsilon(t)
\end{equation}
where $\Re\{\cdot\}$ denotes the real part, and $N$ is the total number of events, taken to be finite herein.  The complex-valued parameter $c_n=|c_n|\re^{\ri \phi_n}$ with $\ri \equiv \sqrt{-1}$ sets the amplitude $|c_n|$ and phase $\phi_n$ of the $n$th event, $t_n$ is its temporal location, and $\rho_n$ sets the event scale. The signal $x_\epsilon(t)$ is a noise process understood to represent all variability not captured by the summation.  The goal of the analysis is to estimate the four signal parameters $|c_n|$, $\phi_n$, $t_n$, and $\rho_n$ for each $n$,  to the extent possible given the noise and interference from other nearby events.  


The representation (\ref{signalmodel}) will be referred to as the {\em element model}, meaning that the signal is believed to be composed of manifestations of the particular function $\psi(t)$, the {\em element function}, which is considered to be known. Note that this model contains a Fourier series plus noise as a special case. Choosing $\psi(t)=e^{\ri t}$, the model becomes $x(t) = \sum_{n=1}^N |c_n| \cos\left(t/\rho_n+\varphi_n\right) +x_\epsilon(t)$, where $\varphi_n\equiv\phi_n-t_n/\rho_n$ is a modified phase that renders the time shift parameter $t_n$ redundant. Because a Fourier series is a very common and powerful representation of signal variability, and because the element model (\ref{signalmodel}) generalizes this to permit the signal to be composed of non-sinusoidal elements, each characterized by four parameters rather than three, this model is likely to be useful for cases in which the Fourier representation is not appropriate.  

\begin{figure}[t!]
\includegraphics[width=1\textwidth]{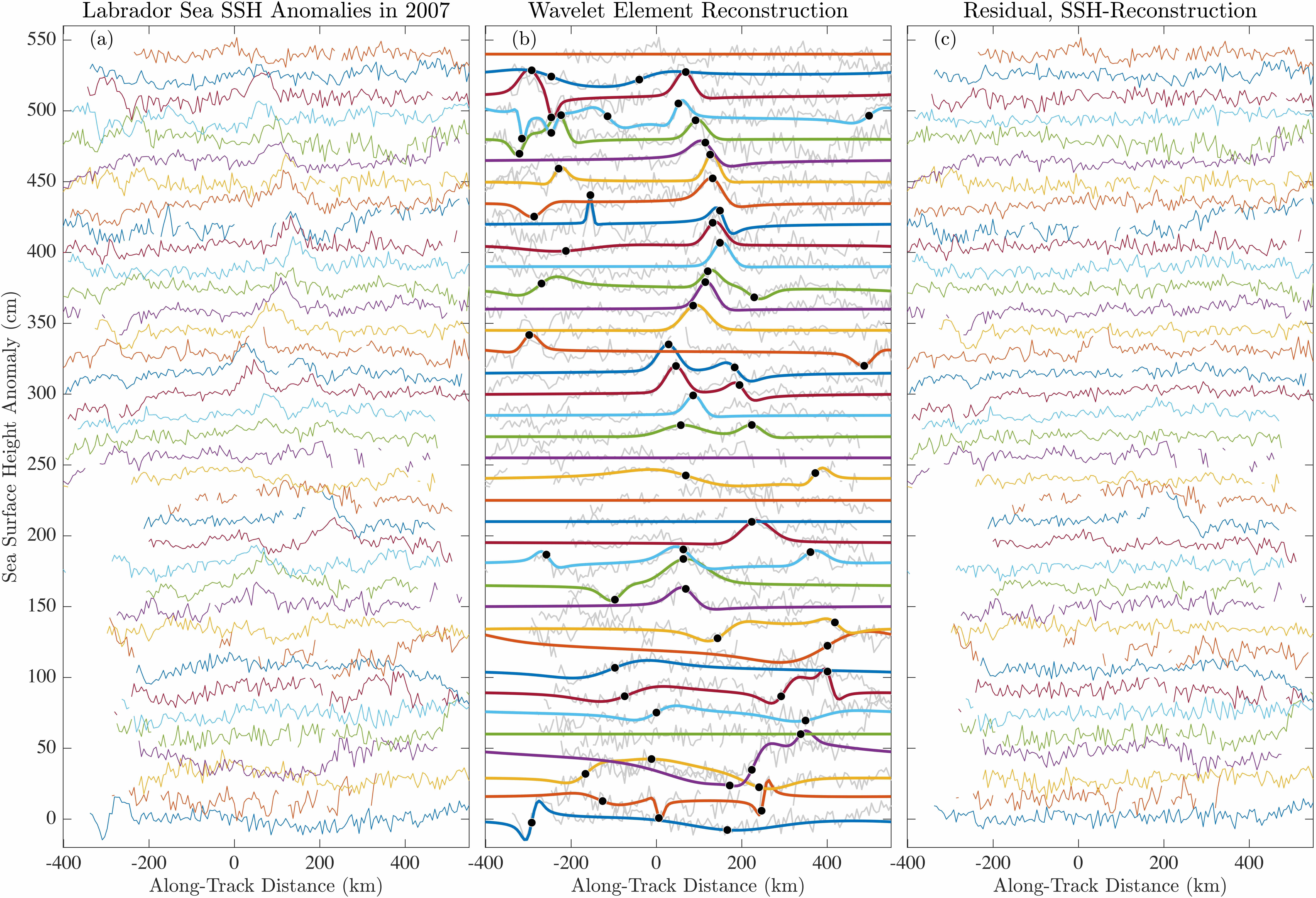}
\caption{The results of applying the element analysis method to a small segment along-track altimeter data.  In (a), along-track sea surface height anomaly measurements are shown from the Labrador Sea, an area of known small-scale eddy activity.  Thirty-seven repeated observations along a single ground-track during the year 2007 are shown, offset from one another by 15~cm with the earliest observations at the bottom.  Reconstructed signals due only to a relatively small number (sixty-seven) of statistically significant and isolated events based on the element model (\ref{morseelementmodel}), determined as described in the text, are shown in (b).  Here the black dots denote the centers of the detected events, while the gray lines repeat the information from (a).  Finally (c) plots the differences between the original signals and the reconstructions, which appear to be largely devoid of meaningful features.   }
\label{impulses-labseathreepanel}
\end{figure}

The element model is directly inspired by continuous wavelet analysis.  If $\psi(t)$ is taken to be a wavelet or integral of a wavelet, (\ref{signalmodel}) can be interpreted as limiting the signal reconstruction to isolated points on the time/scale plane.  The general approach to analyzing a time series that is believe to match the element model has three steps: (i) detecting wavelet transform maxima characterizing the individual events, (ii) determining the level of significance by examining the time/scale distribution of transform maxima arising due entirely to noise, and (iii) ensuring the appropriateness of this model through a criterion for verifying that each event is sufficiently isolated from the others.  Thus, unlike the method of wavelet thresholding \cite{donoho95-itit}, one is not simply looking for statistically significant coefficients, but rather for significant features which are also a good match to the specified element function.  An illustration that this method is able to extract a small number of isolated events from a real-world satellite altimetry dataset, leaving behind apparently unstructed noise, is presented in figure~\ref{impulses-labseathreepanel}, and will be discussed in detail later.  

In order to be a suitable model for a variety of signals, it is essential that the element function $\psi(t)$ be capable of taking on a broad range of forms.  Here the generalized Morse wavelets, or simply the Morse wavelets for brevity, are an attractive choice.  These wavelets were introduced by  \cite{daubechies88-ip}, then examined further by \cite{bayram00-nnsp,olhede02-itsp,olhede03a-prsla,lilly09-itsp,lilly12b-itsp}.  Their fundamental position within the wavelet pantheon is now clear.  Recently it has been shown \cite{lilly12b-itsp} that the Morse wavelets effectively encompass all other types of commonly-used analytic wavelets within a single unified family.  Analytic delta-functions  and complex exponentials are also included as limiting cases.  Therefore, using the generalized Morse wavelets as signal elements provides more flexibility than using all these other types of functions put together.  Furthermore, their simple frequency-domain form means that analytic expressions for key properties may readily be derived \cite{lilly09-itsp}, and thus their dependence on controlling parameters is well understood.  While a Gaussian is the element function of greatest immediate interest to the eddy detection problem, it is not much more difficult to create a general method that can utilize any Morse wavelet as an element function, as is done here. 


The proposed method joins a diverse set of methods already in use in the literature for structure detection and analysis in time series. A straightforward wavelet-based approach is to simply specify a sequence of filtration and / or reconstruction steps that tend to have the effect of isolating structures of interest for a particular problem \cite{thomas04-tac,barthlott07-blm}.  The present method is distinguished from such approaches in that it begins by positing a model (\ref{signalmodel}) for what the signal is actually like.  This allows for the construction of a method for inferring event properties with a small number of adjustable parameters, making the element analysis method highly automatable and scalable. Another, non-wavelet-based approach applies statistical tests to sliding windows of a given length to determine whether they are likely to contain signal structures \cite{kang14-jas,kang15-qjrms}; detected events can then be classified using objective methods.  That approach assumes that typical duration of an event is known, but its form is unknown; in element analysis, we treat the opposite case in which the form is considered to be known but the duration is unknown.   

A sophisticated and powerful approach related to the one proposed here is basis pursuit \cite{chen01-siam}.  In that method, one attempts to find the most compact representation of the signal by considering a variety of complete or overcomplete representations.  Basis pursuit can be implemented as a denoising method by incorporating a penalty function into the optimization, see \S~5 of \cite{chen01-siam}.  Basis pursuit is intended as a general-purpose tool, with the goal of obtaining a compact representation of any structures present in the signal, whatever they may be.  In element analysis, it is assumed that there is a physical motivation for believing that the signal consists of {\em isolated} events of a known form.  The goal is not to reconstruct all signal structure, but rather to infer the properties of those events.  For this specific problem, element analysis has the powerful features of being able to assess the significance of the detected events against the null hypothesis of white or power-law noise, and to reject unsuitable events.  Thus the assumptions and objectives of element analysis are different from those of basis pursuit and other existing structure-detection methods.  The method developed here therefore complements those already existing in the literature.

The structure of the paper is as follows. Essential background on wavelet analysis and the Morse wavelets is presented in \S~\ref{section:background}.  The basic idea of element analysis is introduced in \S~\ref{section:element}.  The means for assessing statistical significance and the degree of isolation are created in \S~\ref{section:significance}. The application to the data shown in figure~\ref{impulses-labseathreepanel} is discussed in \S~\ref{section:application}, and the paper concludes with a discussion.  All software related to this paper is distributed as a part of a freely available toolbox of Matlab functions, called \texttt{jLab}, available at the author's website,  \url{http://www.jmlilly.net}.  Descriptions of relevant routines from this toolbox are given in Appendix~A.  

\section{Background}\label{section:background}

This section presents relevant background on wavelet analysis using the generalized Morse wavelets. This involves briefly reviewing key material from the literature, especially \cite{olhede02-itsp,olhede03a-prsla,lilly09-itsp,lilly12b-itsp}, together with additional details when necessary.  The definition of the continuous wavelet transform is reviewed in \S~\ref{section:background:transform}, while the essential properties of the generalized Morse wavelets are discussed in \S~\ref{subsection:properties}.  An interpretation of the meaning behind maxima of the wavelet transform with the inverse scale normalization employed here is given in \S~\ref{section:optimization}; this is important, as it identifies the optimization principle on which the element analysis is based.  

\subsection{The continuous wavelet transform}\label{section:background:transform}

In this paper we will consider a time series to be built up from members of a two-parameter family of functions termed {\em generalized Morse wavelets} \cite{olhede02-itsp,lilly09-itsp,lilly12b-itsp}. The Morse wavelets, represented as $\psi_{\beta,\gamma}(t)$, are defined in the frequency domain for $\beta\ge 0$ and $\gamma>0$ as, respectively, 
\begin{equation}\label{waveletdef}
\psi_{\beta,\gamma}(t) =\frac{1}{2\pi } \int_{-\infty}^{\infty} \Psi_{\beta,\gamma}(\omega) \re^{\ri \omega t}\,\rd \omega, \quad\quad
\Psi_{\beta,\gamma}(\omega) \equiv a_{\beta,\gamma}\, \omega^\beta \re^{-\omega^\gamma}\times \left\{ \begin{array}{cc} 1 & \omega > 0 \\1/2 & \omega = 0 \\
0 & \omega < 0 \end{array}\right.
\end{equation}
where $\omega$ is angular or radian frequency, and $a_{\beta,\gamma}$ is a real-valued normalizing constant chosen as
\begin{equation}\label{Adef}
a_{\beta,\gamma}\equiv 2\left(\frac{\re\gamma}{\beta}\right)^{\beta/\gamma}
\end{equation}
in which the ``\re'' appearing in the numerator is Euler's number, $\re\approx2.71828$.  The parameter $\beta$, called the {\em order}, controls the low-frequency behavior, while $\gamma$, called the {\em family}, controls the high-frequency decay.  Differentiating $\Psi_{\beta,\gamma}(\omega)$ with respect to $\omega$, one finds that the Morse wavelets obtain their maximum value at the frequency \begin{equation}\label{peakfrequency}\omega_{\beta,\gamma}\equiv (\beta/\gamma)^{1/\gamma}\end{equation} which is known as the {\em peak frequency}. The choice of $a_{\beta,\gamma}$ in (\ref{Adef}) sets the maximum value of the frequency-domain wavelet to $\Psi_{\beta,\gamma}(\omega_{\beta,\gamma})=2$, for reasons to be seen subsequently. 

Functions having no support on negative frequencies, such as the Morse wavelets, are said to be {\em analytic}. Analyticity implies that the wavelets $\psi_{\beta,\gamma}(t)$ must be complex-valued, because the contribution to $\psi_{\beta,\gamma}(t)$ from each complex-valued exponential $\re^{\ri \omega t}$ in (\ref{waveletdef}) cannot be canceled by those at other frequencies.  This means the analytic wavelets are naturally grouped into even or cosine-like and odd or sine-like pairs, allowing them to naturally capture phase variability.

The wavelet transform of a square-integrable signal $x(t)$ with respect to the wavelet $\psi_{\beta,\gamma}(t)$ is defined in the time domain, or the frequency domain, respectively as
\begin{equation}\label{wavetrans}
w_{\beta,\gamma}(\tau,s)
\equiv \int_{-\infty}^{\infty} \frac{1}{s}\,
\psi_{\beta,\gamma}^*\left(\frac{t-\tau}{s}\right)\, x(t) \,\rd t=
\frac{1}{2\pi} \int_{-\infty}^{\infty} \re^{\ri\omega\tau} 
\Psi_{\beta,\gamma}^*\!\left(s\omega\right) X(\omega) \,\rd \omega
\end{equation}
where $X(\omega)$ is the Fourier transform of $x(t)$, with $x(t)=\frac{1}{2\pi}\int_{-\infty}^{\infty}  \re^{\ri\omega t}X(\omega) \rd \omega$, and where the asterisk denotes the complex conjugate. The time-domain expression is the inner product\footnote{ Given two square-integrable functions $f(t)$ and $g(t)$, their Hilbert space inner product is defined as $ \int_{-\infty}^{\infty} f(t) g^*(t) \, \rd t$. } between the signal $x(t)$ and shifted, rescaled versions of the wavelet.  The frequency-domain form is found by inserting the Fourier representations of $x(t)$ and $\psi_{\beta,\gamma}(t)$, then  using $\int_{-\infty}^{\infty} \re^{\ri\omega t} \rd t = 2\pi \delta(\omega)$ where $\delta(\omega)$ is the Dirac delta function, or from Plancherel's formula.  The scale variable $s$ specifies a stretching or compression of the wavelet in time.  The rescaled frequency-domain wavelet $\Psi_{\beta,\gamma}(s\omega)$ obtains a maximum at $\omega_s\equiv\omega_{\beta,\gamma}/s$, referred to here as the {\em scale frequency}.

Note that for convenience herein write the time series of interest as $x(t)$, as if it were observed in continuous time.  In reality, this is not the case, and the time series $x(t)$ is only available as the discrete sequence $x_n\equiv x(\Delta n)$ where $\Delta$ is the sampling interval.  We will discuss discrete effects only when necessary, for example, when discussing numerical implementation. In practice, the discrete effects may be neglected provided we choose the scale $s$ sufficiently large compared to~$\Delta$.

In the above, we have chosen to normalize the time-domain wavelets with $1/s$ as opposed to the more common $1/\sqrt{s}$. The $1/\sqrt{s}$ normalization guarantees that the wavelet maintains constant energy, since $s^{-1}\int_{-\infty}^\infty\left|\psi_{\beta,\gamma}(t/s)\right|^2 \rd t=\int_{-\infty}^\infty\left|\psi_{\beta,\gamma}(t)\right|^2 \rd t$. Thus this normalization is appropriate if one wishes for the modulus-squared wavelet transform to reflect the {\em energy} of the analyzed signal $x(t)$. However, we find it is generally more useful to describe time-localized signals by their {\em amplitude}, and for this the $1/s $ normalization is more appropriate. To see this, we note that compressing or stretching the signal $x(t)$ in time by some factor $\rho$ as in $x(t/\rho)$, but without modifying the signal amplitude, rescales the wavelet transform as
\begin{equation}
\int_{-\infty}^{\infty} \frac{1}{s}\,
\psi_{\beta,\gamma}^*\left(\frac{t-\tau}{s}\right)\, x(t/\rho) \,\rd t=\int_{-\infty}^{\infty} \frac{1}{s/\rho}\,
\psi_{\beta,\gamma}^*\left(\frac{t-\tau/\rho}{s/\rho}\right)\, x(t) \,\rd t=w_{\beta,\gamma}(\tau/\rho,s/\rho)
\end{equation}
as one finds from a change of variables. Thus, rescaling time in the input signal as $x(t/\rho)$ rescales both the time and the scale of the wavelet transform, but without changing its magnitude. The transform values of the amplitude-rescaled signal $c x(t/\rho)$ then reflect the value of $c$, independent of the choice of temporal rescaling $\rho$, a desirable result that is not true with the $1/\sqrt{s}$ normalization. A special case of this result is that the peak magnitude of the wavelet transform of a sinusoid $c \cos\left( \omega_o t\right)$ always takes on the same value regardless of the frequency $\omega_o$. Because of the choice of $a_{\beta,\gamma}$ in (\ref{Adef}), the maximum magnitude of the wavelet transform of this sinusoid obtains a value of $|c|$, which occurs at scale frequency $\omega_s=\omega_o$ or scale $s=\omega_{\beta,\gamma}/\omega_o$. 

The zeroth-order functions $\psi_{0,\gamma}(t)$, with $\beta=0$, require special comment.  These functions are well defined by (\ref{waveletdef}), but are technically not wavelets because wavelets are zero mean by definition. The time-mean value of $\psi_{\beta,\gamma}(t)$ is $\int_{-\infty}^\infty \psi_{\beta,\gamma}(t) \rd t =\Psi_{\beta,\gamma}(0)$, which from (\ref{waveletdef}) is seen to vanish for $\beta>0$ but not for $\beta=0$.  We will therefore refer to $\psi_{\beta,\gamma}(t)$ defined by (\ref{waveletdef}) for any $\beta\ge0$ and positive $\gamma$ as Morse {\em functions} rather than {\em wavelets}, whereas the Morse {\em wavelets} strictly occur for $\beta>0$.  The amplitude coefficient $a_{\beta,\gamma}$ given by (\ref{Adef}) is of the form $0^0$ at $\beta=0$, which by mathematical convention is taken to equal unity.  This gives  $a_{0,\gamma}=2$, consistent with the limiting value of $a_{\beta,\gamma}$ as $\beta$ tends to zero, as is readily shown.  The zeroth-order Morse functions are therefore seen to be one-sided bandpass filters of the form $\Psi_{0,\gamma}(\omega)=2\re^{-\omega^\gamma}$. For these zeroth-order functions $\psi_{0,\gamma}(t)$, we also  need a different way of assigning a reference frequency, since the peak frequency $\omega_{\beta,\gamma}\equiv (\beta/\gamma)^{1/\gamma}$ vanishes in this case.  Instead, we define $\omega_{0,\gamma}$ as the half-power point, i.e. the frequency at which $\Psi_{0,\gamma}(\omega)=2\re^{-\omega^\gamma}$ is equal to half of its maximum value of $\Psi_{0,\gamma}(0)=2$.  Solving $\Psi_{0,\gamma}\left(\omega_{0,\gamma}\right)=1$ then leads to $\omega_{0,\gamma}\equiv\sqrt[\gamma]{\ln(2)}$.

\subsection{Properties of Morse wavelets}\label{subsection:properties}

\begin{figure}[t!]
\begin{center}
\includegraphics[height=17.5cm,angle=-90]{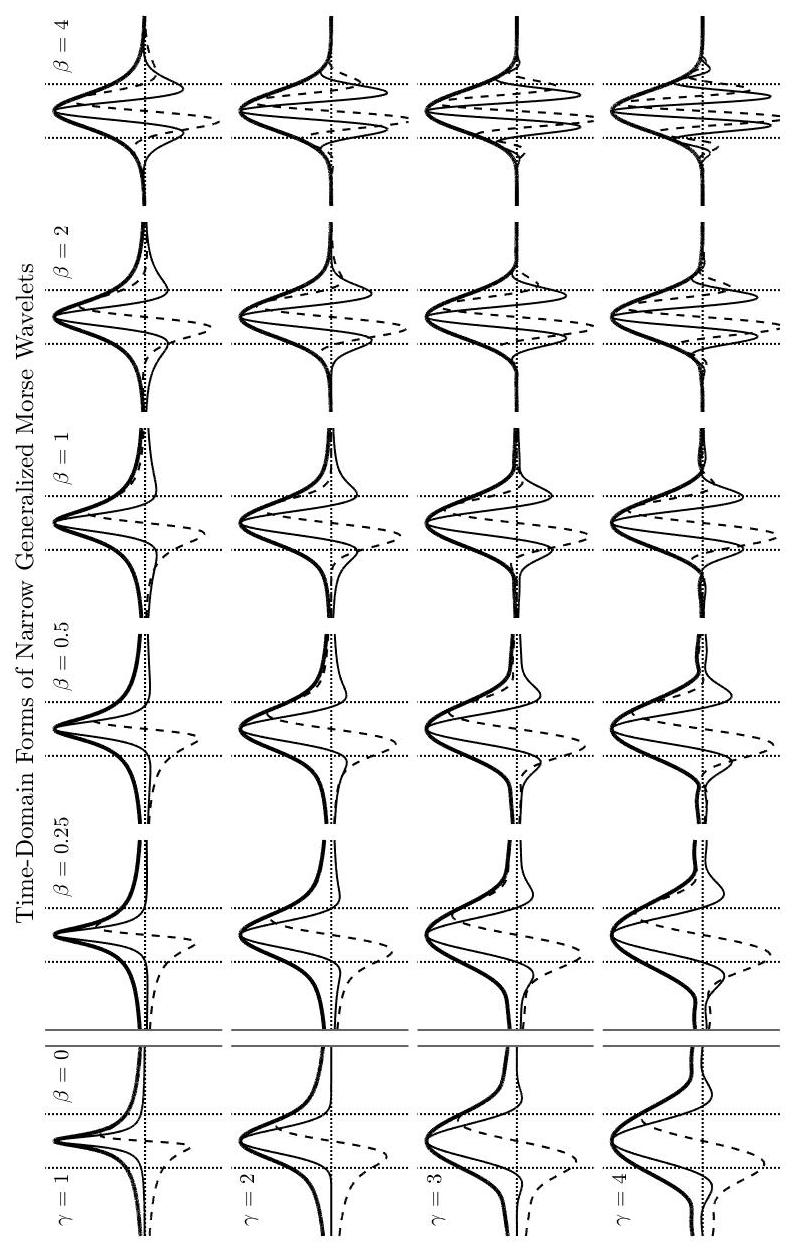}
\end{center}
\caption{ Examples of time-localized signals of the Morse wavelet form. Each row corresponds to a particular value of $\gamma$, and each column to a particular value of $\beta$, as indicated. The real parts, imaginary parts, and absolute values are shown as solid, dashed, and heavy solid lines, respectively. The zero value is marked by the horizontal dotted lines, while the dotted vertical lines mark the time $t=\pm \frac{1}{2}L_{\beta,\gamma}$, defined in (\ref{Ldef}).  The signals in the $\beta=0$ class, to the left of the double solid line, are not classified as wavelets since they are not zero mean.  For these $\beta=0$ functions, $L_{\beta,\gamma}$ is not defined, so the vertical dotted lines mark the comparable interval of $\sqrt{2}$ times the period corresponding to the half-power frequency of the function's Fourier transform, $\pm\sqrt{2}\pi/\omega_{0,\gamma}$, see the last paragraph in \S~\ref{section:background:transform}.  For small $\beta$ and $\gamma$, the signals have an impulsive rather than oscillatory character, becoming more oscillatory toward the lower right as $\beta$ and $\gamma$ increase.}
\label{timedomainwavelets}
\end{figure}

The Morse wavelets can present a wide range of time-domain forms, as shown in figure~\ref{timedomainwavelets} for a variety of values of $\beta$ and $\gamma$. The functions become more oscillatory as one moves across columns, as $\beta$ increases, and also moving down rows as $\gamma$ increases. As these parameters decrease, the functions become increasingly localized in the time domain, appearing more as isolated events or impulses rather than as oscillations. Increasing $\beta$ with fixed $\gamma$ appears to pack more oscillations into the same envelope, whereas increasing $\gamma$ with fixed $\beta$ additionally modifies the function shape, with the function modulus curves becoming less strongly concentrated about its center.  In fact, incrementing $\beta$ by one is essentially equivalent to performing a time derivative, because
\begin{equation}\label{betaderiv}
\psi_{\beta+1,\gamma}(t) = -\ri \frac{ a_{\beta+1,\gamma}}{ a_{\beta,\gamma}}\frac{\rd}{\rd t} \psi_{\beta,\gamma}(t) 
\end{equation}
as can be seen directly from (\ref{waveletdef}).  Thus all  wavelets with $\beta>0$ in the same $\gamma$ family can be generated by repeatedly differentiating (or fractionally differentiating) the zeroth-order generalized Morse functions $\psi_{0,\gamma}(t)$, which are shown separated from the others in the left-hand column of figure~\ref{timedomainwavelets}.  Varying $\gamma$, on the other hand, leads to qualitatively different families.  The most familiar of these is $\gamma=2$, for which the zeroth-order function consists of the analytic part of a Gaussian, with derivatives of this analytic Gaussian occurring for higher-order $\beta$.  For more details on the roles of $\beta$ and $\gamma$ in shaping the wavelet properties, see \cite{lilly09-itsp} and \cite{lilly12b-itsp}.

In addition to the peak frequency $\omega_{\beta,\gamma}$, a second fundamental quantity is a nondimensional measure of the wavelet's time-domain width, denoted $P_{\beta,\gamma}$.  From the definition
\begin{equation}\label{pdef}
P_{\beta,\gamma}^2\equiv \omega_{\beta,\gamma}^2 \frac{ \int_{-\infty}^\infty t^2\,\re^{-\ri\omega_{\beta,\gamma} t}\psi_{\beta,\gamma}(t) \,\rd t}{ \int_{-\infty}^\infty \re^{-\ri\omega_{\beta,\gamma} t}\psi_{\beta,\gamma}(t) \,\rd t}=-\omega_{\beta,\gamma}^2\frac{\Psi_{\beta,\gamma}''(\omega_{\beta,\gamma})}{\Psi_{\beta,\gamma}(\omega_{\beta,\gamma})}
=\beta\gamma
\end{equation}
one sees that $P_{\beta,\gamma}$ is the square root of the second moment of the wavelet, after demodulation by its own peak frequency.  The first equality follows from the Fourier transform $\int_{-\infty}^\infty \re^{-\ri\omega t}\psi_{\beta,\gamma}(t) \,\rd t =\Psi_{\beta,\gamma}(\omega)$, and the fact that $P_{\beta,\gamma}^2$ evaluates to $P_{\beta,\gamma}^2=\beta\gamma$ may be be verified directly, or see \S~III-B of \cite{lilly09-itsp}. Because it is the product of a time-domain width and the wavelet's peak frequency $\omega_{\beta,\gamma}$, $P_{\beta,\gamma}$ could be called the {\em time-bandcenter  product}. The third expression in (\ref{pdef}) suggests that $P_{\beta,\gamma}$ could also be interpreted as a nondimensional {\em inverse bandwidth}, see~\cite{lilly12b-itsp}.  

A dimensional measure of the wavelet's time-domain width will also be utilized.  Note that $P_{\beta,\gamma}/\omega_{\beta,\gamma}$ is a measure of the time-domain half-width of the $s=1$ or ``mother'' wavelet.  Thus we may introduce a measure of the duration of the scale $s$ wavelet, termed the wavelet {\em footprint}, as
\begin{equation}\label{Ldef}
L_{\beta,\gamma}(s)\equiv 2\sqrt{2} \,\frac{P_{\beta,\gamma}}{\omega_s}=  2\sqrt{2} \,\frac{P_{\beta,\gamma}}{\omega_{\beta,\gamma}} s.  
\end{equation}
Through numerical calculation, we find a window of this width typically captures $\approx 95$\% of the total wavelet energy.  As discussed in Appendix~B, the wavelet footprint is closely related to a more familiar quantity, the wavelet's time-domain standard deviation.  

A more complete description of the wavelet properties is provided by the wavelet moments [\S~III-A of \cite{lilly09-itsp}], which will be utilized in several mathematical derivations herein.  The relevant aspects of the wavelet moments are discussed in \S~S1 of the supplementary text.    
 
\subsection{Optimization principle}\label{section:optimization}

In this section we examine the optimization principle on which the $1/s$ or amplitude-normalized wavelet transform is based. To do so, we first examine the more familiar $1/\sqrt{s}$ or energy normalization.  Let us say that we attempt to fit a rescaled and shifted version of a wavelet $\psi_{\beta,\gamma}(t)$ to the real-valued time series $x(t)$ by minimizing the total error 
\begin{equation}\label{expandederror}
\epsilon_{\beta,\gamma}(c,\tau,s) = \int_{-\infty}^\infty \left|x(t) - \Re\left\{\frac{c}{\sqrt{s}}  \psi_{\beta,\gamma}\left(\frac{t-\tau}{s}\right)\right\}\right|^2 \rd t
\end{equation}
and we therefore seek the coefficients $c$, $\tau$, and $s$ that minimize this error. The choice of $c$ that minimizes the error will be denoted $c_{\beta,\gamma}(\tau,s)$.  Setting the partial derivatives of $\epsilon_{\beta,\gamma}(c,\tau,s)$ with respect to the real and imaginary parts of $c$ equal to zero, one finds
\begin{equation}
c_{\beta,\gamma}(\tau,s)  = 2\calE_{\beta,\gamma}^{-1}\int_{-\infty}^\infty  \frac{1}{\sqrt{s}}  \psi_{\beta,\gamma}^*\left(\frac{t-\tau}{s}\right) x(t)  \rd t 
\end{equation}
where $\calE_{\beta,\gamma}\equiv \int_{-\infty}^\infty \left|\psi_{\beta,\gamma}(t)\right|^2 \rd t$ is the energy of the scale $s=1$ or ``mother'' wavelet.  This states that the best fit coefficient at each time and each scale is proportional to the continuous wavelet transform with a $1/\sqrt{s}$ or energy normalization.  Inserting this expression into (\ref{expandederror}) leads to
 \begin{equation}
\epsilon_{\beta,\gamma}(c_{\beta,\gamma},\tau,s) =  \int_{-\infty}^\infty \left|x(t)\right|^2 \rd t -\frac{1}{2}\calE_{\beta,\gamma}\left|c_{\beta,\gamma}(\tau,s) \right|^2
 \end{equation}
and since the first term is constant, the error is minimized for that choice of time offset $\tau$ and scale parameter $s$ that maximize the squared modulus of the energy-normalized wavelet transform, in other words, for $(\tau,s)$ being a maximum point of the wavelet transform modulus.

Thus the maxima points of the energy-normalized wavelet transform give the {\em local best fits}---in the sense of minimizing the time-integrated error---between the observed signal $x(t)$ and time-shifted, rescaled versions of the wavelet.  While this might appear a compelling argument to use this normalization, when we carry out the analysis described herein using the energy normalization on real-world data, the results are poor.  The reason is that the energy-normalized wavelet transform is overly influenced by variability at adjacent times.  In fact, attempting to explain as much variability as possible using a single wavelet is not a suitable principle for analyzing time series containing multiple, potentially interacting events.  Because longer wavelets can capture more energy, the transform has a tendency to achieve a maximum when it is long enough to span several nearby events; but the objective here is to detect the events individually.

The quantity $\frac{1}{2}\calE_{\beta,\gamma}\left|c_{\beta,\gamma}(\tau,s) \right|^2$ is that portion of the total signal energy that can be explained by a single wavelet located at time $\tau$ and scale $s$; note that it has units of energy, like $\int_{-\infty}^\infty \left|x(t)\right|^2 \rd t$.  The related quantity
\begin{equation}
\frac{1}{4} \calE^2\frac{1}{s}\left|c_{\beta,\gamma}(\tau,s) \right|^2 = \left|\int_{-\infty}^\infty  \frac{1}{s}  \psi_{\beta,\gamma}^*\left(\frac{t-\tau}{s}\right) x(t)  \rd t \right|^2 = \left| w_{\beta,\gamma}(\tau,s)\right|^2
\end{equation}
is therefore proportional to the {\em energy density} in a time interval of duration $s$, or the {\em power} captured by a wavelet located at a particular time/scale point.  This is the same as the wavelet transform with an amplitude or $1/s$ normalization.  Therefore, maxima points of the amplitude-normalized wavelet transform identify the time offsets $\tau$, scaling factors $s$, and complex-valued coefficients $c$ that maximize the energy {\em density} over an interval proportional to their own duration.  Thus the $1/s$-normalized wavelet transform is based on the principle of {\em optimizing power}.

\section{Element analysis}\label{section:element}

In this section, element analysis using the Morse wavelets is developed. It is shown that if the element function $\psi(t)$ in (\ref{signalmodel}) is chosen to be a Morse function, then analyzing the signal $x(t)$ with any Morse wavelet in the same $\gamma$ family leads to a straightforward way of inferring the event properties.  Firstly, in \S~\ref{section:element:morse}, the use of the Morse functions as signal elements is introduced, and transform maxima points are defined.  In \S~\ref{section:element:transforms}, it is shown that the wavelet transform of a Morse function with another Morse wavelet can itself be expressed as a modified Morse wavelet.  This fact lets us derive, in \S~\ref{section:element:values}, a simple expression for the entire wavelet transform of a time series represented by the element model.  In that section we also find expressions for the time/scale points at which transform maxima should occur, and the values of those maxima, given the properties of the underlying signal elements.  Thus, properties of observed transform maxima can be inverted to obtain estimates of the element properties, as shown in \S~\ref{section:element:inferring}.  Finally, an illustration of transform maxima in a synthetic dataset is given in \S~\ref{section:element:examples}, the examination of which motivates the development of statistical significance and degree of isolation criteria in the next section.

\subsection{Generalized Morse functions as signal elements}\label{section:element:morse}

Here we propose the use of the Morse functions $\psi_{\beta,\gamma}(t)$ as element functions, leading to a signal model of the form  
\begin{equation}\label{morseelementmodel}
x(t) = \sum_{n=1}^N \Re \left\{ c_n \psi_{\mu,\gamma}\left(\frac{t-t_n}{\rho_n}\right)\right\} +x_\epsilon(t)
\end{equation}
where the properties of the element function are set by $\mu$ and $\gamma$, and where $x_\epsilon(t)$ is a noise process defined subsequently and which is assumed to be zero mean.  The parameter $\mu$ plays the role of $\beta$ in the element function, while $\rho$ plays the role of the scale $s$; we reserve $\beta$ and $s$ to refer later to the  analyzing wavelet. Note that $\mu$, unlike $\beta$, can be equal to zero.  Taking the wavelet transform of $x(t)$ using a $(\beta,\gamma)$ Morse wavelet leads to \begin{equation}\label{transformofelementmodel}
w_{\beta,\gamma}(\tau,s)=\frac{1}{2}\sum_{n=1}^N c_n\int_{-\infty}^{\infty} \frac{1}{s}
\psi_{\beta,\gamma}^*\left(\frac{t-\tau}{s}\right)\psi_{\mu,\gamma}\left(\frac{t-t_n}{\rho_n}\right)\,\rd t+\varepsilon_{\beta,\gamma}(\tau,s)
\end{equation}
where $\varepsilon_{\beta,\gamma}(\tau,s)$ denotes the wavelet transform of the noise process $x_\epsilon(t)$.  Here, we have written  the real part in (\ref{morseelementmodel}) as $\Re\left\{z\right\}=\frac{1}{2}[z+z^*]$, then noted that the wavelet transform of the anti-analytic function $\psi_{\mu,\gamma}^*(t)$ with the analytic wavelet $\psi_{\beta,\gamma}(t)$ vanishes identically, as can readily be seen from the frequency-domain form of the wavelet transform in (\ref{wavetrans}).

We define {\em transform maxima points} as time/scale locations $(\hat\tau,\hat s)$ at which the wavelet transform modulus takes on a local maximum, that is, a point at which 
\begin{equation}
\frac{\partial}{\partial \tau}\left|w_{\beta,\gamma}(\tau,s)\right| =\frac{\partial}{\partial s}\left|w_{\beta,\gamma}(\tau,s)\right|=0, \quad\quad\frac{\partial^2}{\partial \tau^2}\left|w_{\beta,\gamma}(\tau,s)\right|<0,\quad\frac{\partial^2}{\partial s^2}\left|w_{\beta,\gamma}(\tau,s)\right|<0.\label{maxconditions}
\end{equation}
The basic idea of element analysis is that the values of the wavelet transform at these points can be used to estimate the coefficients $c_n$, scales $\rho_n$, and temporal locations $t_n$ of the $N$ events comprising the signal in the model (\ref{morseelementmodel}). There are three aspects to this analysis.  Firstly we show how in the absence of noise, and assuming the $N$ events are sufficiently well-separated in time and in scale, the event properties $t_n$, $\rho_n$, and $c_n$ may be recovered from the maxima points of the wavelet transform.  Secondly, we examine the wavelet transform of noise, and establish the rate at which ``false positive'' maxima occur due to idealized noise processes.  This leads to the establishment of a threshold cutoff associated with a particular density of spurious maxima points associated with the noise.  Thirdly, we enforce the condition that the remaining, statistically significant maxima points are well-separated using a region-of-influence condition.


\subsection{The Morse transform of another Morse function}\label{section:element:transforms}

The wavelet transform of the $\mu$th-order Morse function $\psi_{\mu,\gamma}(t/\rho)$ with a $\beta$th-order Morse wavelet in the same $\gamma$ family has a simple expression, and is given by
\begin{equation}\label{morseofmorse}
\int_{-\infty}^{\infty} \frac{1}{s}
\psi_{\beta,\gamma}^*\left(\frac{t-\tau}{s}\right)\psi_{\mu,\gamma}\left(\frac{t}{\rho}\right)\,\rd t=\zeta_{\beta,\mu,\gamma}(\tau/\rho,s/\rho)
\end{equation}
where $\zeta_{\beta,\mu,\gamma}(\tau,s)$ is a modified wavelet function defined as 
\begin{equation}\label{simplephi}
\zeta_{\beta,\mu,\gamma}(\tau,s) \equiv
\frac{ a_{\beta,\gamma}\,a_{\mu,\gamma}}{a_{\beta+\mu,\gamma}}\frac{s^{\beta}}{\left(\sqrt[\gamma]{s^\gamma+1}\,\right)^{\beta+\mu+1}}
\,\psi_{\beta+\mu,\gamma}\left(\frac{\tau}{\sqrt[\gamma]{s^\gamma+1}}\right).
\end{equation}
The derivation may be found in \S~S2 of the supplemental text.  The wavelet transform of a Morse function with a Morse wavelet in the same $\gamma$ family is therefore itself expressible as a modified version of a Morse wavelet.  Furthermore, (\ref{morseofmorse}) shows that taking the wavelet transform of the rescaled Morse function $\psi_{\mu,\gamma}(t/\rho)$ implies rescaling both the time and the scale of the wavelet transform of the original function $\psi_{\mu,\gamma}(t)$, but without changing the transform amplitude. 

The main feature in (\ref{simplephi}) is the appearance of a wavelet with order $(\beta+\mu)$. Both $\beta$ and $\mu$ correspond to powers of $\omega$ in the frequency domain, which can be combined because the wavelet transform corresponds to a multiplication in the frequency domain. The scale dependence reveals two distinct effects: a more involved dependence of the {\em amplitude} on the scales $s$ and $\rho$ than the usual $1/s$, and more significantly a {\em rescaling} of the time argument of wavelet that is itself a function of the transform scale $s$. To understand the scale dependence of (\ref{simplephi}) in more detail, we examine the large-scale and small-scale limits to find
\begin{equation}\label{simplephilimits}
\zeta_{\beta,\mu,\gamma}(\tau/\rho,s/\rho) \approx \frac{ a_{\beta,\gamma}\,a_{\mu,\gamma}}{a_{\beta+\mu,\gamma}} \times \left\{ \begin{array}{ll} (\rho/s)^{\mu+1} \,\psi_{\beta+\mu,\gamma}(\tau/s) & \quad \quad s\gg \rho\\(s/\rho)^{\beta} \,\psi_{\beta+\mu,\gamma}(\tau/\rho) & \quad\quad s\ll \rho 
\end{array}\right.
\end{equation}
which has an illuminating interpretation. When $s\gg\rho$, the analyzing wavelet $\psi_{\beta,\gamma}(t/s)$ is much broader than the signal element $\psi_{\mu,\gamma}(t/\rho)$, and consequently the wavelet smooths the signal, spreading the transform out over the wavelet scale $s$. However, when $s\ll\rho$, the wavelet is much {\em narrower} than the signal, and the transform scale remains fixed at the scale $\rho$ of the analysed signal, simply decaying in magnitude as $s$ decreases further. 

\subsection{Transform values at transform maxima}\label{section:element:values}

The $\zeta_{\beta,\mu,\gamma}(\tau,s)$ function allows us to determine the values of the wavelet transform at maxima points, and relate these to the properties of the signal elements.  In terms of $\zeta_{\beta,\mu,\gamma}(\tau,s)$, we have
\begin{equation}
w_{\beta,\gamma}(\tau,s) = \frac{1}{2}\sum_{n=1}^N c_n \zeta_{\beta,\mu,\gamma}\left(\frac{\tau-t_n}{\rho_n},\frac{s}{\rho_n}\right)+ \varepsilon_{\beta,\gamma}(\tau,s)\label{compactexpression}
\end{equation} 
as a compact expression for the wavelet transform of the element model presented in (\ref{transformofelementmodel}).  The expected value of the squared modulus of the wavelet transform is then approximately given by
\begin{equation}\label{squaredelements}
\E\left\{\left|w_{\beta,\gamma}(\tau,s)\right|^2\right\} \approx \frac{1}{4} \sum_{n=1}^N |c_n|^2 \left| \zeta_{\beta,\mu,\gamma}\left(\frac{\tau-t_n}{\rho_n},\frac{s}{\rho_n}\right)\right|^2  + \E\left\{\left|\varepsilon_{\beta,\gamma}(\tau,s)  \right|^2\right\}
\end{equation}
if one neglects the interactions between different terms in the summation; here $\E\{\cdot\}$ denotes the statistical expectation. The cross-terms between the noise and the wavelet transforms of the element functions vanish in expectation on account of the zero mean assumption.  We assume that the events are sufficiently well separated such that (\ref{squaredelements}) is a good approximation within a certain time/scale region surrounding each transform maxima, as discussed in detail  in \S~\ref{subsection:region}.

Under the approximation (\ref{squaredelements}), if the function $|\zeta_{\beta,\mu,\gamma}(\tau,s)|$ decays monotonically from its maximum value, then if noise is neglected there will be exactly one transform maxima for each of the $N$ events.  There are two caveats to this.  Firstly, low-level maxima may arise due to the event-event interactions that are neglected by (\ref{squaredelements}), as will be illustrated later in a supplemental figure. Secondly, for some extreme parameter choices with large values of $\gamma$ and small values of $\beta$, the wavelet modulus may not decay monotonically in time from the wavelet center.  In those wavelets, one sometimes sees small sidelobe maxima, see e.g. the $(\beta,\gamma)=(1/2,4)$  wavelet in figure~\ref{timedomainwavelets}.  If $|\zeta_{\beta,\mu,\gamma}(\tau,s)|$ does not decay monotonically, then one would expect to see minor maxima on the flanks of each primary maxima associated with the $N$ signal elements.  Both of these issues lead to weak spurious maxima.  For well-separated signal elements, these will either be below the noise level, or may be easily rejected with an amplitude cutoff. The approximation (\ref{squaredelements}) allows us to focus on the primary maxima that describe the events within the context of the element model.  
 


We now find the scale locations and transform values associated with the maximum points of the wavelet transform of a Morse function.   The maximum value of $\left|\zeta_{\beta,\mu,\gamma}(\tau/\rho,s/\rho)\right|$ for all times and all scales is found to occur at time $\tau=0$ and normalized scale $s/\rho=\tilde s_{\beta,\mu,\gamma}^{\,\max}$, with a value of
\begin{equation}\label{maxs}
\zeta_{\beta,\mu,\gamma}^{\max}\equiv\zeta_{\beta,\mu,\gamma}\left(0,\tilde s_{\beta,\mu,\gamma}^{\,\max}\right)
,\quad\quad\quad\tilde s_{\beta,\mu,\gamma}^{\,\max}\equiv
\left(\frac{\beta}{\mu+1}\right)^{1/\gamma}.
\end{equation}
To see this, we note that maximum of $|\zeta_{\beta,\mu,\gamma}(\tau/\rho,s/\rho)|$ with respect to variations in time occurs at $\tau=0$.  At this time $\zeta_{\beta,\mu,\gamma}(0,s/\rho)$ takes on the real and positive value 
\begin{equation}\label{phimax}
\zeta_{\beta,\mu,\gamma}(0,\tilde s) = \frac{ a_{\beta,\gamma} a_{\mu,\gamma}}{2\pi\gamma}\, \Gamma\left(\frac{\beta+\mu+1}{\gamma}\right) \,\frac{\tilde s^{\beta}}{\left(\sqrt[\gamma]{\tilde s^\gamma+1}\right)^{\beta+\mu+1}}
\end{equation}
introducing the normalized scale $\tilde s \equiv s/\rho$.  This follows by combining (\ref{simplephi}) with the expression for $\psi_{\beta,\gamma}(0)$ given in \S~S1 of the supplementary text. Differentiating $\zeta_{\beta,\mu,\gamma}(0,\tilde s)$ with respect to $\tilde s$, one finds that this quantity obtains a global maximum for any $\tilde s$ at the value $\tilde s = \tilde s_{\beta,\mu,\gamma}^{\,\max}$ given by (\ref{maxs}).


The maximum value of the Morse wavelet transform of another Morse function is found, inserting (\ref{maxs}) into (\ref{phimax}), to be given by
\begin{equation}\label{zetamaxdef}
\zeta_{\beta,\mu,\gamma}^{\max} = \frac{ a_{\beta,\gamma} a_{\mu,\gamma}}{2\pi\gamma}\, \Gamma\left(\frac{\beta+\mu+1}{\gamma}\right)\vartheta_{\beta,\mu,\gamma}
\end{equation}
where we have defined for future reference the scale weighting function
\begin{equation}\label{simplephitildelevelset}
\vartheta_{\beta,\mu,\gamma}\equiv\frac{\left(\tilde s_{\beta,\mu,\gamma}^{\,\max}\right)^{\beta}}{\left[\left(\tilde s_{\beta,\mu,\gamma}^{\,\max}\right)^\gamma+1\right]^{(\beta+\mu+1)/\gamma}}= \frac{\left(\frac{\beta}{\mu+1}\right)^{\beta/\gamma}}{\left(\frac{\beta}{\mu+1}+1\right)^{(\beta+\mu+1)/\gamma}}.
\end{equation}
The maximum value $\zeta_{\beta,\mu,\gamma}^{\max}$ is seen to be independent of the scale $\rho$ of the transformed function.

\subsection{Inferring element properties from maxima points}\label{section:element:inferring}

Under the assumption that the noise process $x_\epsilon(t)$ vanishes, and subject to the caveats regarding spurious minor maxima discussed above, there will be one maximum point of the transform modulus associated with each of the $N$ elements.  The $n$th maximum point will be located at time $t_n$ and scale $s_n=\rho_n \tilde s_{\beta,\mu,\gamma}^{\,\max}$, and from (\ref{compactexpression})  we find that the wavelet transform at this point is
\begin{equation}
w_{\beta,\gamma}\left(t_n,s_n\right) = \frac{1}{2} c_n \zeta_{\beta,\mu,\gamma}^{\max}.
\end{equation}
Thus, one may work backwards from the set of observed time/scale maxima, at locations denoted by $(\hat\tau_n,\hat s_n)$  determined from the transform as in (\ref{maxconditions}), to infer or estimate the element properties $(t_n,\rho_n,c_n)$.  Defining $\hat w_n\equiv w_{\beta,\gamma}(\hat \tau_n,\hat s_n)$ as the transform at the $n$th observed maxima, we have 
\begin{equation}\label{inferences}
\hat t_n=\hat \tau_n \quad\quad\quad\hat \rho_n = \frac{\hat s_n}{\tilde s_{\beta,\mu,\gamma}^{\,\max}}\quad\quad\quad
\hat c_n=2\frac{\hat w_n}{\zeta_{\beta,\mu,\gamma}^{\max}} 
\end{equation}
where the hatted quantities $\hat t_n$, $\hat \rho_n$, and $\hat c_n$ indicate {\em inferences} for the values of element properties based on the transform maximum points. Thus the properties of the events can be read off directly from the maxima of the wavelet transform, provided the element function is considered as known.  

Because the implementation used here refers to wavelets by their frequencies rather than their scales, the expression (\ref{inferences}) mapping $\hat \rho_n$ into $\hat s_n$ needs to be modified.  The scale frequency characterizing scale $s$ of the transform is $\omega_s=\omega_{\beta,\gamma}/s$, while $\omega_{\rho}=\omega_{\mu,\gamma}/\rho$ is the  frequency at which the scale~$\rho$ element function obtains a maximum value.  Substituting $s=\omega_{\beta,\gamma}/\omega_s$ and $\rho=\omega_{\mu,\gamma}/\omega_\rho$ into $\hat \rho_n = \hat s_n/\tilde s_{\beta,\mu,\gamma}^{\,\max}$ from (\ref{inferences}) for the scale location of a maximum point, one finds
\begin{equation}\label{rhotos}
 \omega_{\hat\rho_n}=\omega_{\hat s_n}\frac{\omega_{\mu,\gamma}}{\omega_{\beta,\gamma}}\tilde s_{\beta,\mu,\gamma}^{\,\max}=\omega_{\hat s_n}\frac{\omega_{\mu,\gamma}}{\omega_{\beta,\gamma}}\left(\frac{\beta}{\mu+1}\right)^{1/\gamma}
\end{equation}
as the relationship between the frequency band $\omega_{\hat s_n}$ of the $n$th observed maximum of the wavelet transform, and the inferred frequency $\omega_{\hat\rho_n}$ characterizing the corresponding element function.

\begin{figure}[t!]
\includegraphics[width=\textwidth]{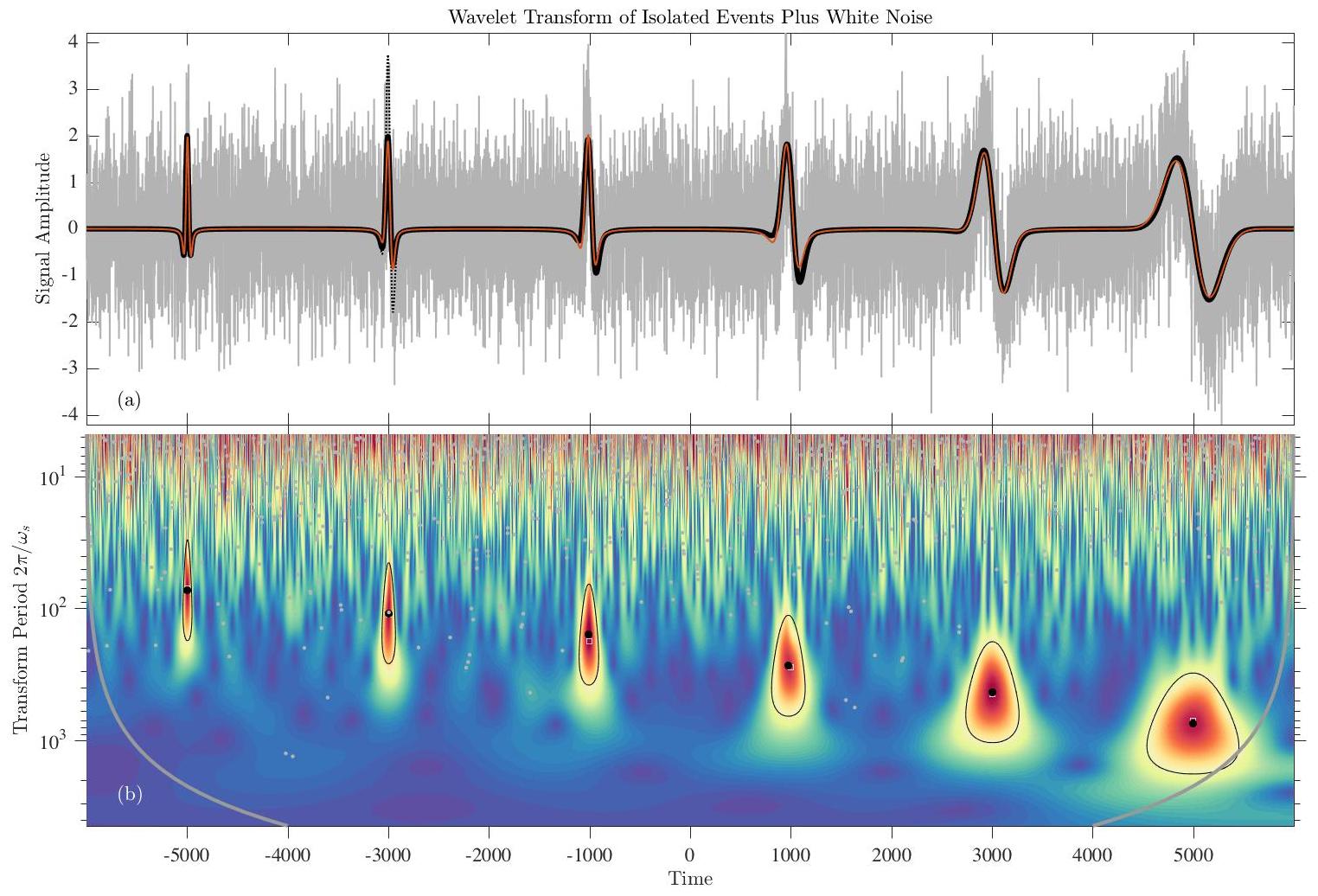}
\caption{An illustration of signal reconstruction using element analysis applied to a synthetic signal in noise.  A 12000 point signal (a) consisting of a series of six impulsive events, constructed from a $\psi_{1,2}(t)$ wavelet as described in the text, is shown as the heavy black line.  This signal is added to a realization of unit-variance Gaussian white noise, resulting in the gray line.  The wavelet transform of this noisy signal with a $\psi_{2,2}(t)$ wavelet is shown in (b), in which the $y$-axis shows the transform period $2\pi/\omega_s$ on a logarithmic scale.  Black circles mark the locations of six statistically significant and isolated maxima, identified as described in the text, while gray dots mark the locations of all other transform maxima.  The locations of the six significant maxima in the absence of noise are shown with gray squares, but these are usually not visible because they are overlapped by the black circles.  The heavy gray curves in (b) denote locations that are contaminated by edge effects, defined as time-scale locations within an interval $L_{\beta,\gamma}(s)/2=\sqrt{2}P_{\beta,\gamma}\omega_s$ from the beginning or the end of the time series.  Black lines delineate the $\lambda=1/2$ region of influence around each maximum, as defined in \S~\ref{subsection:region}.  In (a), the red line shows the reconstruction based on statistically the six significant and isolated maxima, which is seen to be virtually identical to the original.  The dotted line shows a reconstruction that does not taken into account the isolation criteria for the maxima, resulting in misfit near the second event; this line is elsewhere obscured by the red line. }
\label{impulses-morsetrain-noisy}
\end{figure}

\subsection{Examples of transform maxima}\label{section:element:examples}

An illustration is presented in figure~\ref{impulses-morsetrain-noisy}. Here the original signal shown in the upper panel is of the form (\ref{morseelementmodel}), with $N=6$ events using first-order Gaussian wavelet $\psi_{1,2}(t)$ as the element function, a 200 point interval between successive events, and with amplitude and scale coefficients given shortly.  Note that the sampling interval in this example is set to unity. We see from figure~\ref{impulses-morsetrain-noisy}a that the events vary from left to right from an even, or cosine-like form, to an odd or negative sine-like form.  The scale increases from left to right, while the maximum excursion decreases somewhat.   To this signal, a realization of unit variance Gaussian white noise has been added.  The modulus of the wavelet transform of the resulting noisy signal with a $\psi_{2,2}(t)$ wavelet is shown in the lower panel. It is seen that the element function scale appears to be increasing at a linear rate along the logarithmic scale axis, while the peak value of the transform modulus appears constant.

The scale frequencies $\omega_{\rho_n}$ for the six events shown here are chosen to vary over a decade from $\omega_{\rho_1}=2\pi/100$ to $\omega_{\rho_6}=2\pi/1000$, with a logarithmic spacing such that $\log_{10}(\omega_{\rho_{n}}/\omega_{\rho_{n+1}})=0.2$ for all $n$.  The coefficient phase, defined as $\phi_n$ in $c_n=|c_n|\re^{\ri\phi_n}$, is set to $\phi_n=(n-1)\pi/10$, and varies from zero to $\pi/2$ as $n$ varies from 1 to 6.  The coefficient amplitude is chosen such that $|c_n\psi_{1,2}(t)|=2$ for all $n$, and is given by $|c_n|=5.39$, see \S~S1 in the supplemental text.  The apparent slight decrease in amplitude in figure~\ref{impulses-morsetrain-noisy}a is actually a consequence of the changing phase.  


The gray dots together with the black circles denote maxima points of the wavelet transform, determined using a numerical approximation to the conditions (\ref{maxconditions}) described in Appendix~A.  On account of the noise, there are many such maxima.  However, six of these maxima, those denoted by the black circles, are found to be both highly statistically significant as well as isolated from one another and from the time series edges, using criteria to be developed in what follows.  From the transform values at these points, we form estimates $(\hat t_n,\hat\rho_n,\hat c_n)$ of the element properties using (\ref{inferences}). Then using these inferred properties, the original signal is reconstructed by inserting the hatted values into (\ref{morseelementmodel}) with $\psi_{1,2}(t)$ as the element function.  The resulting reconstruction, shown as the red curve, is virtually identical to the original signal, despite the fact that the original signal is almost totally obscured by the noise.  Similar results are obtained for the same signal added to a realization of unit-variance red noise, computed by cumulatively summing discrete Gaussian white noise, as presented in the supplemental figure~S1.

\section{Significance and isolation}\label{section:significance}

In this section we determine two criteria that must be applied to the transform maxima in order to identify meaningful events within the context of the element model.  The first is a measure of statistical significance, and the second is a measure of isolation from other transform maxima.  It will be assumed that the noise has a power-law spectrum, a form that encompasses both white noise and fractional Brownian motion.  The expected value of the modulus-squared wavelet transform---or {\em wavelet spectrum}---of power-law noise is derived in \S~\ref{section:significance:noise}.  The next step is to find the distribution of transform maxima due entirely to the presence of noise, as this will allow the significance of detected events to be determined.  This is accomplished in \S~\ref{section:noisesim} with the help of a Monte Carlo method that sidesteps the need to take the wavelet transform of noise realizations, and that instead allows the covariance properties of the wavelet spectrum to be simulated directly. Properties of transform maxima arising from noise are then examined in \S~\ref{subsection:noisedist} and used to establish statistical significance.  The final step in the algorithm is to determine whether the detected events are sufficiently isolated from one another such that the element model appears to be suitable.  This is addressed in \S~\ref{subsection:region} with the identification of new type of region associated with the Morse wavelet transform of another Morse function, referred to as the {\em region of influence}.

\subsection{The wavelet transform of noise}\label{section:significance:noise}

Now we consider the wavelet transform of the noise $x_\epsilon(t)$, which is assumed to be zero mean, stationary, and Gaussian.  Due to the assumption of stationarity, the noise process has a Cram\'{e}r spectral representation of the form
\begin{equation}
\label{xepsilonrepresentation}
x_\epsilon(t)= \frac{1}{2\pi}\int_{-\infty}^{\infty}  e^{\ri\omega t} \,\rd X_\epsilon(\omega)
\end{equation}
where $X_\epsilon(\omega)$ is an orthogonal increment process, i.e. $\E\left\{\rd X_\epsilon(\omega) \,\rd   X_\epsilon^*(\nu)  \right\}$ vanishes unless  $\omega=\nu$. The spectrum of $x_\epsilon(t)$ is defined in terms of its orthogonal increment process as 
\begin{equation}\label{noisespectrumdefinition}
S_\epsilon(\omega) \delta(\omega-\nu)  \rd\omega\rd\nu\equiv \frac{1}{2\pi} \E\left\{\rd X_\epsilon(\omega) \,\rd   X_\epsilon^*(\nu)  \right\}
\end{equation} 
with $\delta(\omega)$ again being the Dirac delta function. Using the spectral representation of $x_\epsilon(t)$, its wavelet transform is given by
\begin{equation}
\label{epsilonrepresentation}
\varepsilon_{\beta,\gamma}(\tau,s) \equiv  \int_{-\infty}^{\infty} \frac{1}{s}\,
  \psi_{\beta,\gamma}^*\left(\frac{t-\tau}{s}\right)\,x_\epsilon (t)\,\rd t=\frac{1}{2\pi}\int_{0}^{\infty} e^{\ri\omega \tau}\Psi_{\beta,\gamma}(s\omega) \,\rd X_\epsilon(\omega)
\end{equation}
and the expected value of the squared modulus of this quantity is found to be
\begin{equation}
\E\left\{\left|\varepsilon_{\beta,\gamma}(\tau,s)  \right|^2\right\} = 
\frac{1}{(2\pi)^2}\int_{0}^{\infty}\int_{0}^{\infty} \Psi_{\beta,\gamma}(s\omega) \,\Psi_{\beta,\gamma}^*(s\nu) e^{i(\omega-\nu) t}\, \E\left\{\rd X_\epsilon(\omega) \,\rd   X_\epsilon^*(\nu)  \right\}.
\end{equation}
Using the definition of the noise Fourier spectrum (\ref{noisespectrumdefinition}), this becomes
\begin{equation}\label{expectednoisetransform}
\E\left\{\left|\varepsilon_{\beta,\gamma}(\tau,s)  \right|^2\right\} =  \frac{1}{2\pi}\int_{0}^{\infty} \left|\Psi_{\beta,\gamma}(s\omega)\right|^2 S_\epsilon(\omega) \rd \omega
\end{equation}
which is independent of time $\tau$, and is found by projecting the Fourier spectrum onto rescaled versions of the modulus-squared Fourier-domain wavelet.  For brevity, we will refer to this expected modulus-squared wavelet transform simply as the {\em wavelet spectrum} of the noise.

Herein we will consider both Gaussian white noise as well as Gaussian red noise having a power-law spectrum.  The latter is important because many time series, geophysical time series especially, have signals embedded in red background noise [e.g. \cite{mann96-cc}].
The red noise case will be considered first.  Assume that a stationary process has the power-law spectrum 
\begin{equation}\label{powerlaw}
S_\epsilon(\omega) = \frac{A^2}{\omega^{2\alpha}}
\end{equation}
with $A$ setting the spectral level and $\alpha$ controlling the spectral slope.  The corresponding wavelet spectrum is found to be
\begin{equation}\label{sigmavss}
\sigma_{\alpha,\beta,\gamma}^2(s)\equiv\E\left\{\left|\varepsilon_{\beta,\gamma}(\tau,s)  \right|^2\right\} =  A^2 f_{\alpha,\beta,\gamma} \,s^{2\alpha-1}
=A^2 f_{\alpha,\beta,\gamma}\left[\frac{\omega_{\beta,\gamma}}{\omega_s}\right]^{2\alpha-1}
\end{equation}
provided $\beta>\alpha-\frac{1}{2}$, as shown in \S~S3 in the supplemental text.  In the above we have introduced the function 
\begin{equation}\label{simplified}
f_{\alpha,\beta,\gamma}\equiv \frac{1}{2\pi}\int_{0}^{\infty}\omega^{-2\alpha}\,\Psi_{\beta,\gamma}^2(\omega) \rd \omega=\frac{a_{\beta,\gamma}^2}{2\pi \gamma} \frac{\Gamma\left(\frac{2\beta-2\alpha+1}{\gamma}\right)}{2^{(2\beta-2\alpha+1)/\gamma}} 
\end{equation}
where the final expression follows from the definition of the gamma function, or see \S~S3.  Thus the wavelet spectrum depends on scale frequency as $\omega_s^{-2\alpha+1}$, which differs from the Fourier spectrum by a factor of $\omega_s$.  This difference can be traced to our choice of the $1/s$ normalization, which we have argued is more appropriate for interpreting the values of transform maxima.  For this reason, the ``wavelet spectrum'' with the $1/s$ normalization should be understood as not being  strictly comparable to the Fourier spectrum.  

The power-law spectrum (\ref{powerlaw}) corresponds for $\frac{1}{2}<\alpha<\frac{3}{2}$ to a random process $x_\epsilon(t)$ consisting of fractional Brownian motion \cite{mandelbrot68-siam}, see \cite{lilly17-npg} and references therein.  Although fractional Brownian motion is itself not stationary, as we have assumed above, a {\em damped} version of fractional Brownian motion known as the Mat\'{e}rn process is stationary \cite{lilly17-npg}. The Mat\'{e}rn process has a spectrum that approximates the power law form (\ref{powerlaw}) for $\omega$ sufficiently greater than zero, as controlled by the damping parameter, and with the slope parameter in the range $1/2<\alpha<\infty$.  Here we will just consider that the noise is stationary and has a spectrum that is equal to or closely approximated by (\ref{powerlaw}) over the frequency range of interest, without specifying the type of the noise process. 


Next we consider the case of Gaussian white noise, which can be considered a special case of the power law spectrum (\ref{powerlaw}) with  $\alpha=0$.  Whereas the Mat\'{e}rn process and fractional Brownian motion are both defined on continuous time, Gaussian white noise is a discrete process and its spectrum is therefore periodized.  With a sampling interval of $\Delta=1$, the noise variance is related to the physically realizable spectrum supported over plus or minus the Nyquist frequency as
\begin{equation}\label{sigepsdef}
\sigma_\epsilon^2 \equiv \E\left\{\left|x_\epsilon(t) \right|^2\right\}= \frac{1}{2\pi}\int_{-\pi}^\pi A^2 \rd \omega = A^2.
\end{equation}
Using this result (\ref{sigmavss}) becomes for $\alpha=0$ 
\begin{equation}\label{sigmaeps}
\sigma_{0,\beta,\gamma}^2(s)=  \sigma_\epsilon^2 f_{0,\beta,\gamma} \frac{1}{s}=\sigma_\epsilon^2 f_{0,\beta,\gamma} \frac{\omega_s}{\omega_{\beta,\gamma}}
\end{equation}
which links the spectral amplitude $A$ to the transform variance for the white noise case.

\begin{figure}[t!]
\includegraphics[width=0.95\textwidth]{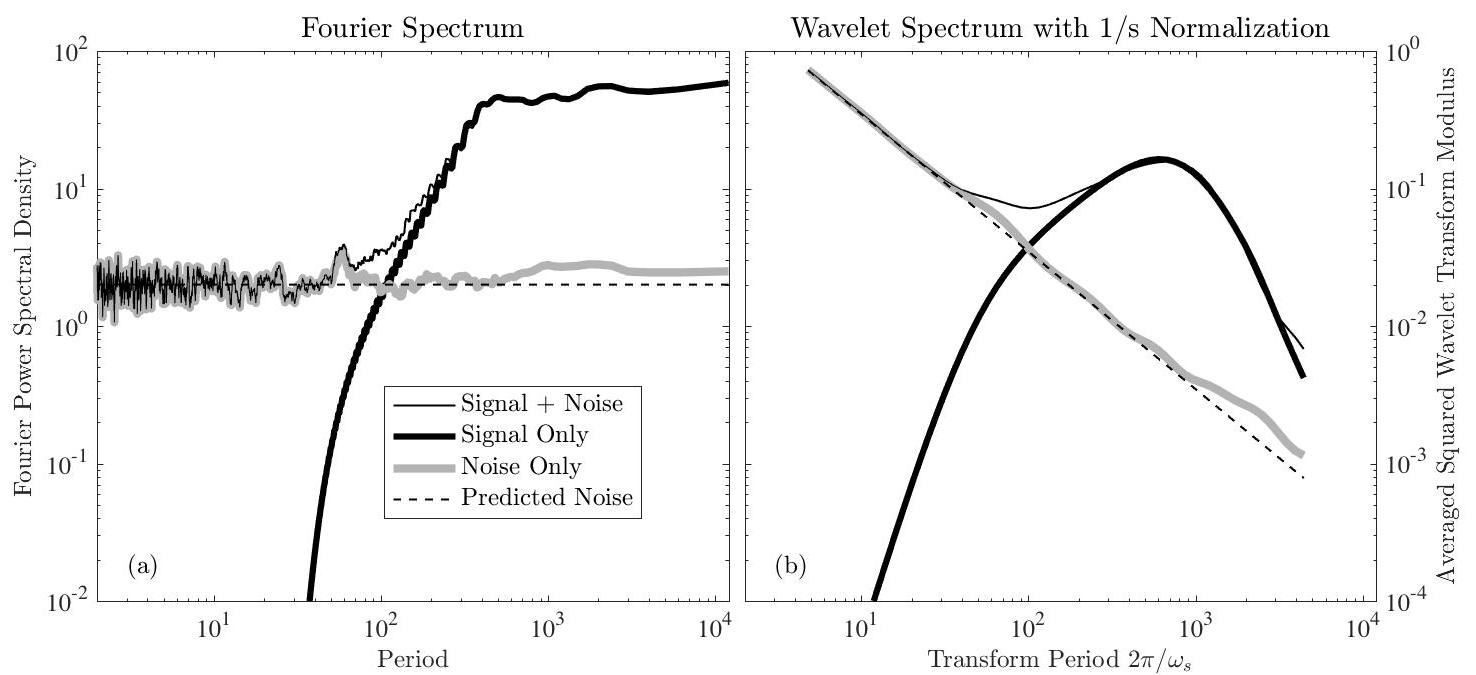}
\caption{The estimated one-sided Fourier spectra (a) and wavelet spectra (b) for the clean and noisy versions of the time series shown in figure~\ref{impulses-morsetrain-noisy}a, together with the spectra of the noise only.  The Fourier spectra in (a) have been computed using the adaptive multitaper method of \cite{thomson82-ieee} with a time-bandwidth product set to 20.  Panel (b) shows the time averages of the magnitude-squared wavelet transforms using  a $\psi_{2,2}(t)$ Morse wavelet, with the heavy black curve being the time average of the square of the transform shown in  figure~\ref{impulses-morsetrain-noisy}b. The $x$-axis, which is the same for both (a) and (b), is presented in terms of period instead of frequency. The dotted lines show the predicted spectral levels for unit variance Gaussian white noise, given by a value of two for the one-sided Fourier spectra in (a) and (\ref{sigmaeps}) for the wavelet spectra in (b).}
\label{impulses-spectra}
\end{figure}


A comparison of the Fourier and wavelet spectra is shown in figure~\ref{impulses-spectra}.  Spectra of three signals are presented: the noisy and original signals from figure~\ref{impulses-morsetrain-noisy}, and their difference which is a time series of unit-variance white noise. In both plots we see that signal dominates noise for periods greater than about 100 data points, whereas noise dominates at smaller scales.  As mentioned earlier, the time-average of the modulus-squared wavelet transform is not an approximation to the Fourier spectrum with the $1/s$ normalization.   In both panels, the dashed line shows the prediction for unit-variance noise.  In the one-sided presentation of the Fourier spectral levels employed here, spectral densities are doubled, so the unit variance signal $x_\epsilon(t)$ has a spectral level of two.  The prediction for the wavelet transform of noise is given by evaluating (\ref{sigmaeps}) with the choices $\sigma_\epsilon=1$, $\beta=2$, and $\gamma=2$.  The realized and predicted noise levels match closely for both the Fourier spectrum and the wavelet transform.

\subsection{Distribution of transform maxima in noise}\label{section:noisesim}

In order to assess the confidence of detected transform maxima, it is necessary to know the rate at which spurious maxima occur due entirely to the background noise.  The distributions of transform maxima in noise can be determined using Monte Carlo simulations, in which one simulates a large time series of power-law noise $x_\epsilon(t)$, takes its wavelet transform $ \varepsilon_{\beta,\gamma}(\tau,s)$, and then searches for transform maxima. This is computationally expensive, particularly because of the need to work with noise time series much longer than the time series of interest in order to obtain stable statistics.  Fortunately the desired statistics can be obtained in a more direct manner.  For Gaussian noise having the power-law spectrum $S_\epsilon(\omega) = A^2 \omega^{-2\alpha}$, the distribution of transform maxima can be determined by simulating a noise vector having the same covariance structure as the wavelet transform at location $(\tau,s)$ and its adjacent four points, as is now shown.  

The task of determining the distribution of transform maxima due to noise may be simplified by recognizing that apart from discretization effects, suitably normalized transform maxima are expected to exhibit a universal distribution across scales.  If at each scale, we normalize transform maxima by the expected root-mean-square magnitude of the wavelet transform of the noise
\begin{equation}
    \widetilde w_n\equiv\hat w_n/\sigma_{\alpha,\beta,\gamma}(\hat s_n)\label{wtilde}
\end{equation}
then the distribution of normalized transform maxima values over a time interval that is the same duration as the scale $s$ wavelet, e.g. as measured by the wavelet footprint $L_{\beta,\gamma}(s)$, should be independent of the scale~$s$.  This conjecture of an approximately universal distribution of transform maxima across scales for a particular choice of $\alpha$, $\beta$, and $\gamma$ will be verified shortly.  If it holds, one would only need to determine the temporal density and amplitude distribution of events at one scale, and then extrapolate to any other scale.



The covariance between the wavelet transform of the noise $\varepsilon_{\beta,\gamma}(\tau,s)$ and itself at another time and another scale is given by the function
\begin{equation}\label{covdef}
\Xi_{\alpha,\beta,\gamma}(u,s,r)  \equiv \E\left\{ \varepsilon_{\beta,\gamma}(\tau,s) \, \varepsilon_{\beta,\gamma}^*(\tau+u,rs) \right\}
\end{equation}
utilizing the fact that  $\varepsilon_{\beta,\gamma}(\tau,s)$ is both zero mean and stationary.  Here $u$ is the time shift between the two versions of $\varepsilon_{\beta,\gamma}(\tau,s)$, while $r$ is the ratio of their scales.  For power-law noise this becomes
\begin{equation}\label{covresult}
\Xi_{\alpha,\beta,\gamma}(u,s,r)  = \frac{\sigma_{\alpha,\beta,\gamma}^2(s)}{f_{\alpha,\beta,\gamma}}\frac{a^2_{\beta,\gamma}}{a_{2\beta-2\alpha,\gamma}}\left[\frac{r^{\beta}}{(1+r^\gamma)^{(2\beta-2\alpha+1)/\gamma}}\right]  \psi_{2\beta-2\alpha,\gamma}^*\!\left(\frac{u}{s\sqrt[\gamma]{1+r^\gamma}}\right)
\end{equation}
as shown in \S~S4 in the supplemental text.  This expression contains three parts: an $s$-dependent coefficient, proportional to the wavelet spectrum of the noise $\sigma_{\alpha,\beta,\gamma}^2(s)$; an $r$-dependent coefficient in square brackets; and a modified version of the $(2\beta-2\alpha,\gamma)$ wavelet containing all the $u$-dependence.   As a check, it is shown in \S~S4 that if $r=1$ and $u=0$, one recovers the wavelet spectrum $\Xi_{\alpha,\beta,\gamma}(0,s,1) =\sigma^2_{\alpha,\beta,\gamma}(s)$, as expected.

Now let $\bx_{\beta,\gamma}(\tau,s)$ be a 5-vector consisting of the noise transform $\varepsilon_{\beta,\gamma}(\tau,s)$ at time $\tau$ and scale~$s$, as well as the noise transform at the four adjacent points on the time/scale plane,
\begin{equation}
\bx_{\beta,\gamma}(\tau,s)\equiv\begin{bmatrix}
 \varepsilon_{\beta,\gamma}(\tau,s)&
 \varepsilon_{\beta,\gamma}(\tau+\Delta,s)&
 \varepsilon_{\beta,\gamma}(\tau-\Delta,s)&
 \varepsilon_{\beta,\gamma}(\tau,rs)&
 \varepsilon_{\beta,\gamma}(\tau,s/r)
\end{bmatrix}^T
\end{equation}
where the superscript ``$T$'' denotes the transpose.  Here $\Delta$ is the sampling interval, and we let the ratio between successive scales,~$r$, take on the value of the scale discretization used in the wavelet transform, see Appendix~C.  The covariance structure of the vector $\bx_{\beta,\gamma}(\tau,s)$, normalized by the local transform variance $\sigma_{\alpha,\beta,\gamma}^2(s)$,  is given by the $5\times 5$ matrix
\begin{equation}
\bm\Sigma_{\alpha,\beta,\gamma}(s) \equiv  \frac{1}{\sigma_{\alpha,\beta,\gamma}^2(s)}\,\E\left\{\bx_{\beta,\gamma}(\tau,s)\,\bx_{\beta,\gamma}^H(\tau,s)\right\}
\end{equation}
which from stationarity is independent of time $\tau$.  For power-law noise, the entries of this matrix can be immediately written down in terms of the transform covariance function $\Xi_{\alpha,\beta,\gamma}(u,s,r)$ as
\begin{multline}
\bm\Sigma_{\alpha,\beta,\gamma}(s) =\frac{1}{\sigma^2_{\alpha,\beta,\gamma}(s)}\times\\
\begin{bmatrix*}[l]
\Xi(0,s,1)&\Xi(1,s,1)&\Xi(-1,s,1) &\Xi(0,s,r)&\Xi(0,s,1/r) \\
\Xi^*(1,s,1)&\Xi(0,s,1)&\Xi(-2,s,1) &\Xi(-1,s,r)&\Xi(-1,s,1/r) \\
\Xi^*(-1,s,1)&\Xi^*(-2,s,1)&\Xi(0,s,1) &\Xi(1,s,r)&\Xi(1,s,1/r) \\
\Xi^*(0,s,r)&\Xi^*(-1,s,r)&\Xi^*(1,s,r) &\Xi(0,rs,1)&\Xi(0,rs,1/r^2) \\
\Xi^*(0,s,1/r)&\Xi^*(-1,s,1/r)&\Xi^*(1,s,1/r) &\Xi^*(0,rs,1/r^2)&\Xi(0,s/r,1) 
\end{bmatrix*}\label{bigSigma}
\end{multline}
in which subscripts have been omitted on  $\Xi_{\alpha,\beta,\gamma}(u,s,r)$  for clarity.  In deriving the above, we have made use of the symmetry $\Xi(u,s,r)=\Xi^*(-u,rs,1/r)$, apparent from the definition (\ref{covdef}), as well as the choice $\Delta=1$.

The distribution of transform maxima due entirely to noise can now be determined as follows.  Decomposing $\bm\Sigma_{\alpha,\beta,\gamma}(s)=\bm L \bm L^H$ using the Cholesky decomposition leads to a lower triangular matrix $\bm L$.  With $\bm\epsilon$ being realizations of a 5-vector containing independent, unit variance, complex-valued Gaussian white noise, we create $\by_{\alpha,\beta,\gamma}(s) \equiv \bm L \bm\epsilon$ at each scale and note that $\E\left\{\by\by^H\right\}=\bm\Sigma_{\alpha,\beta,\gamma}(s)$ by construction.  In other words, $\by_{\alpha,\beta,\gamma}(s)$ has the same covariance structure as we would observe by grouping the wavelet transform of power-law noise at point $(\tau,s)$ with its four neighbors into the vector $\bx_{\beta,\gamma}(\tau,s)$.  The probability that the first element of $\by_{\alpha,\beta,\gamma}(s)$, denoted $y_1$, is greater in magnitude than the other four elements is the same as the probability of there being a transform maxima in $\varepsilon_{\beta,\gamma}(\tau,s)$ at scale $s$.  Similarly, the amplitude distribution of $y_1$ given that it is the largest-magnitude element in  $\by_{\alpha,\beta,\gamma}(s)$  will be the same as the amplitude distribution of normalized maxima values $\widetilde w_n$ in $\varepsilon_{\beta,\gamma}(\tau,s)$.  Thus rather than simulating noise and taking its wavelet transform, we can simulate $\by_{\alpha,\beta,\gamma}(s)$ directly by creating realizations of a 5-vector of noise and then performing a matrix multiplication---a considerable simplification.


\subsection{Simulations of noise distributions}\label{subsection:noisedist}

\begin{figure}[t!]
\includegraphics[width=\textwidth]{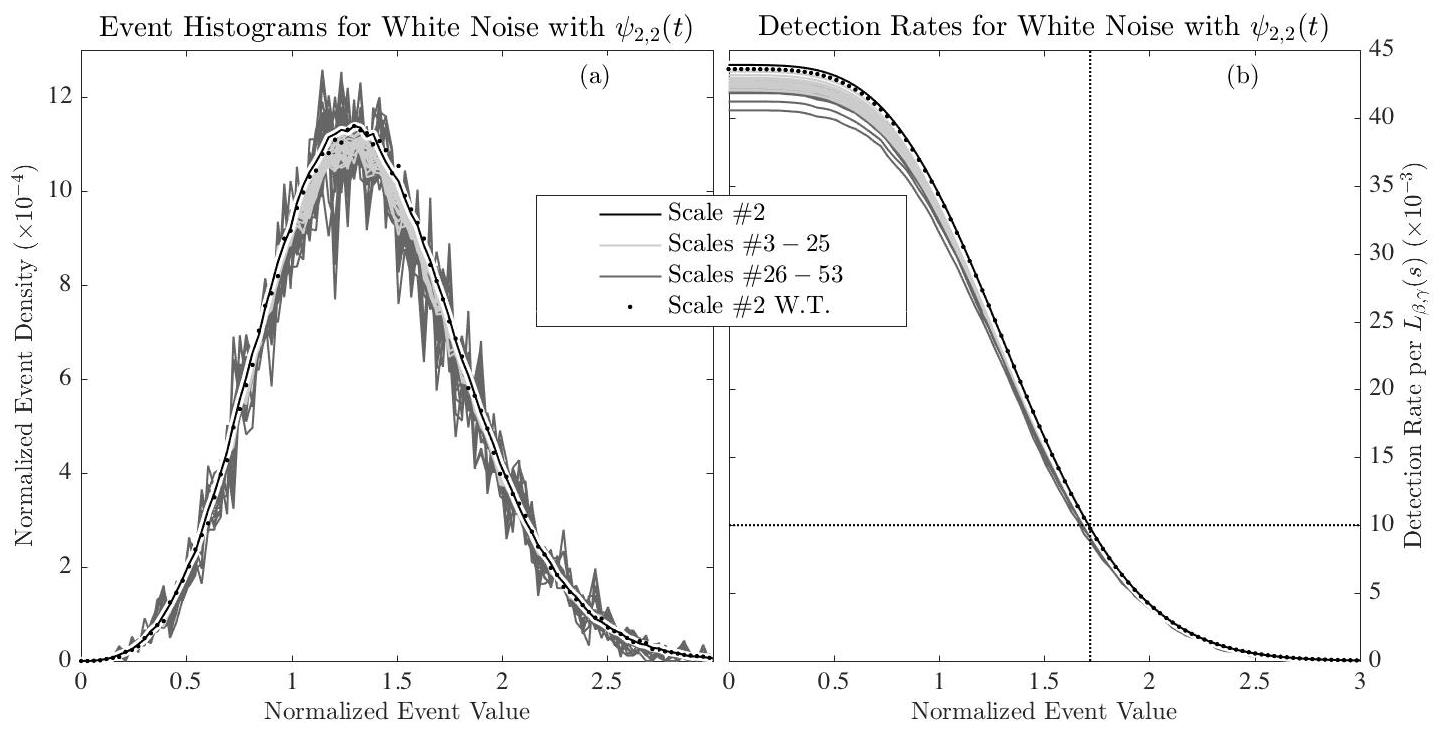}
\caption{Normalized histograms (a) and false detection rates (b) for transform maxima within unit-variance Gaussian white noise transformed with a $\psi_{2,2}(t)$ wavelet, as was used in figure~\ref{impulses-morsetrain-noisy}, computed by simulating the $\by_{\alpha,\beta,\gamma}(s)$ vector as described in the text.  Panel (a) shows the distribution of the normalized event magnitude $|\widetilde w_n|=|\hat w_n|/\sigma_{\alpha,\beta,\gamma}(s)$ with a bin size of about 0.03, and with the curves for different scales normalized such that they sum to the total number of events expected in a time series of length $L_{\alpha,\beta,\gamma}(s)$.  In (b), the cumulatively summed distributions corresponding to these curves are shown, but summed from large values to small in order to indicate the rate of an event occurring that is larger than a particular value.  As an example, the horizontal line indicates a rate of one event per $100L_{\alpha,\beta,\gamma}(s)$, which occurs at a value of $|\widetilde w_n|\approx 1.7$, marked by the vertical line.  Fifty-two curves are shown in each panel, corresponding to all 59 scale frequency bands used in figure~\ref{impulses-morsetrain-noisy}, apart from the first and the last where no maxima can be detected.  These plots are created by simulating $10^{4}M$ realizations of $\by_{\alpha,\beta,\gamma}(s)$, where $M=12000$ is length of the time series in figure~\ref{impulses-morsetrain-noisy}.  The dots are a comparison from explicitly taking the wavelet transform of noise at the three smallest scale bands and searching for transform maxima within the second band, using a times series of length 2000~$M$; see text for details. }
\label{impulses-noisedist-22}
\end{figure}

An example of this approach to simulating the distribution of maxima in noise is shown in figure~\ref{impulses-noisedist-22} for a white noise time series analyzed with a $\psi_{2,2}(t)$ wavelet, as in the example of figure~\ref{impulses-morsetrain-noisy}. For each of the 59 scale bands shown in that transform, we compute $\bm\Sigma_{\alpha,\beta,\gamma}(s)$ from (\ref{bigSigma}) with $\beta=2$ and $\gamma=2$, and with $\alpha=0$ corresponding to a white noise process $x_\epsilon(t)$.  We then simulate $12000\times 10^{4}$ realizations of $\by_{\alpha,\beta,\gamma}(s)$ at each scale, following the steps in the previous paragraph.  The histogram of the amplitudes of $|y_1|$ when it is the largest element in the vector simulates the histogram of the normalized amplitudes of transform maxima $|\widetilde w_n|=|\hat w_n|/\sigma_{\alpha,\beta,\gamma}(s)$.  These histograms are shown in figure~\ref{impulses-noisedist-22}a for 57 scale bands, excluding the first and the last, as transform maxima cannot be numerically detected there.  The histogram is computed in 100 bins linearly spaced between zero and three.  The $y$-value of each curve is normalized such that the curve sums to the total number of maxima points detected in a time series one wavelet footprint $L_{\beta,\gamma}(s)$ in duration.  In other words, both the $x$-axis and the $y$-axis have been normalized in accordance with the universal distribution proposed in the second paragraph in the previous section.  

The curve for the second scale band, shown as the black line, is approximately Gaussian in shape, with a mean value of about 1.36 indicating that a typical transform maxima has a non-normalized value  $|\hat w_n|$ somewhat larger in magnitude than the expected transform amplitude $\sigma_{\alpha,\beta,\gamma}(s)$, an intuitive result.  However, it is rare to find values of $|\widetilde w_n|$ exceeding 2, with only about 10\% of the transform maxima having larger magnitudes.  A slight tendency for positive skewness is apparent, as may be expected due to the fact that $|\widetilde w_n|$ is non-negative.  

Examining the curves from all the scales, we see that normalizing the amplitudes by $\sigma_{\alpha,\beta,\gamma}(s)$ and the densities by $L_{\alpha,\beta,\gamma}(s)$ has indeed virtually collapsed all the curves together, in agreement with the proposed universal distribution.  The most significant difference is that as one proceeds to larger scales, a higher degree of scatter is observed.  This occurs because within $\by_{\alpha,\beta,\gamma}(s)$, the first component $y_1$ becomes increasingly correlated with the other four components as $s$ increases; thus the effective sample size of a fixed-length simulation decreases, increasing the variance.  Within an intermediate band of scales, from bands 3 to 25, there is a tendency for the central peak to decrease slightly as scale increases, although this tendency does not appear to continue indefinitely.  The conjecture of a universal distribution therefore appears to be a close but not quite perfect approximation.  The minor dependence of the maxima statistics on $s$ are attributed to discretization effects, which are correctly captured by the simulations based on $\bm\Sigma_{\alpha,\beta,\gamma}(s)$. 

For comparison, the distribution of transform maxima points for the second scale band are also computed by explicitly taking the wavelet transform of a noise time series.  A real-valued Gaussian white noise time series $x_\epsilon(t)$ of length $12000\times 2000$ is transformed with the $\psi_{2,2}(t)$ wavelet using only the first three scale bands, or the three smallest-scale wavelets, used in  figure~\ref{impulses-morsetrain-noisy}b.  Transform maxima may then be identified in the second band, and their normalized distributions are plotted in figure~\ref{impulses-noisedist-22}a as black dots.  The agreement with the  calculation based on simulating $\by_{\alpha,\beta,\gamma}(s)$ in the second scale band, shown as the black line, is excellent.   Normalized distribution curves for other choices of $\beta$ and $\gamma$, which are not shown, are similar in form to those shown in figure~\ref{impulses-noisedist-22}a, and are generally roughly Gaussian in shape with a slight positive skewness.

In figure~\ref{impulses-noisedist-22}b, the cumulative distributions associated with these histogram curves are shown, but summed in the reverse direction from large values to small values.  This quantity is known in the literature as the {\em survival function} or {\em complementary cumulative distribution function}; in the context of this analysis it will be shown to indicate a false detection rate.  The curves in figure~\ref{impulses-noisedist-22}b  give the rate at which transform maxima {\em larger} than a particular value occur.  The highest value for all the curves, near zero amplitude, indicates than a transform maxima with {\em any} amplitude occurs at a rate of about 0.040--0.045 events over one wavelet footprint $L_{\alpha,\beta,\gamma}(s)$, or one maxima every 22--25 footprints.  The horizontal line marks a rate of 0.01 events per $L_{\alpha,\beta,\gamma}(s)$ or one maxima every 100 footprints, and is found to be the rate at which events larger in magnitude than about 1.7 $\sigma_{\alpha,\beta,\gamma}(s)$ occur.  Such curves can be used to set an amplitude cutoff for a tolerable false detection rate.  Most of the differences between the rate curves occur for small-amplitude transform maxima; for amplitudes greater than about unity, the curves are all virtually indistinguishable.  

\begin{figure}[t!]
\includegraphics[width=1\textwidth]{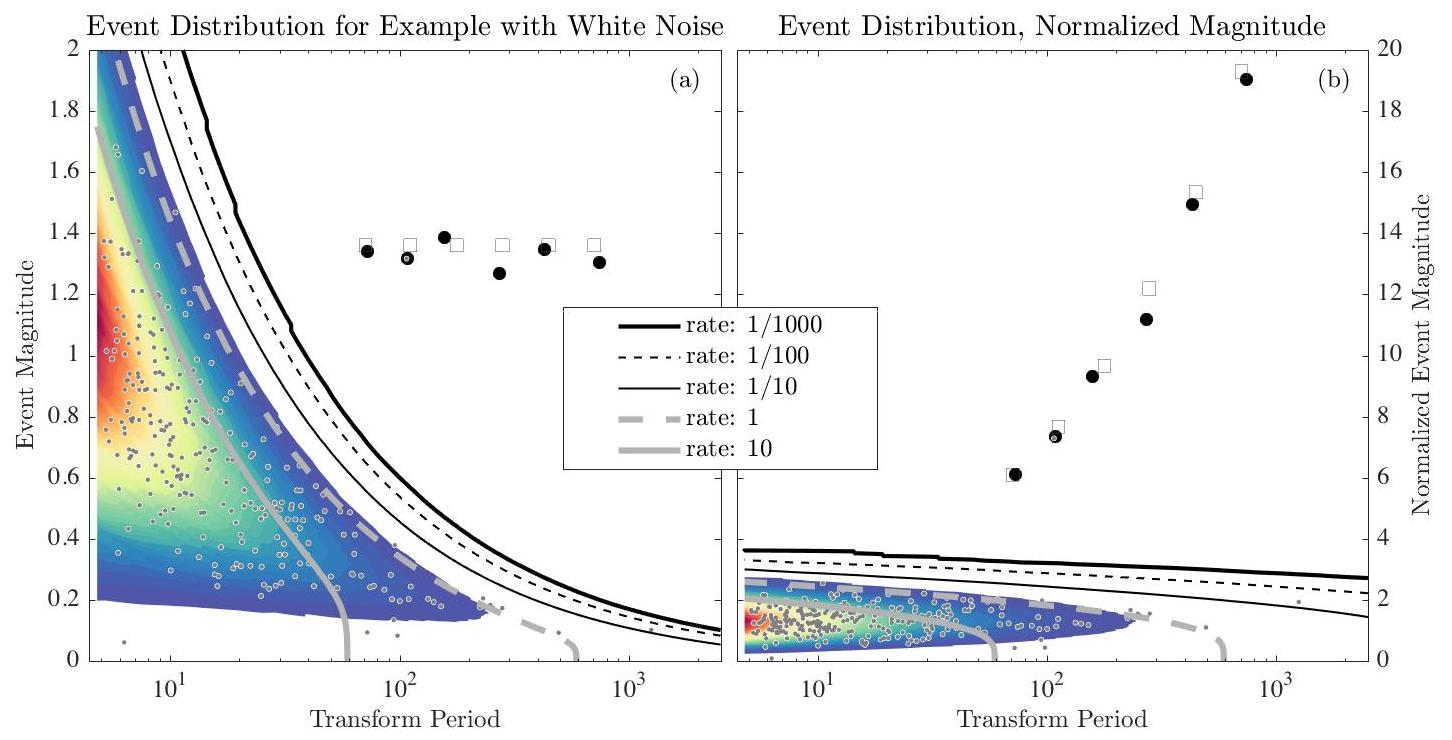}
\caption{The distribution of transform maxima for the example shown in figure~\ref{impulses-morsetrain-noisy}, shown in two different ways.  In (a), gray dots mark the periods $2\pi/\omega_{\hat s_n}$ of all detected maxima plotted against their magnitudes $|\hat w_n|$.  In (b), the $y$-axis now shows $|\widetilde w_n|$ as defined in (\ref{wtilde}), the transform maxima value normalized by the expected noise level at each scale.  Note that the $x$-axis in both panels is the scale frequency expressed as a period, $2\pi/\omega_s$. In both panels, the black dots show the scale locations and magnitudes of six significant and isolated maxima of the noisy wavelet transform, as shown earlier in figure~\ref{impulses-morsetrain-noisy}, while the gray squares show the actual scale locations and magnitudes of the six events in the noise-free signal.  The colored shading shows the density of transform maxima observed in a large noise simulation using the $\by_{\alpha,\beta,\gamma}(s)$-vector method, as described in the text.  The individual contours show curves of false detection rate inferred from the colored shading.  For example, the heavy black curve labeled ``rate: 1/1000'' means that at each scale level, a transform maxima of this amplitude or larger is expected to occur only once per 1000 realizations of a time series of this length ($M=12000$).  The colored shading corresponds to the density seen in seen in figure~\ref{impulses-noisedist-22}a, while the contours correspond to detection rates seen in figure~\ref{impulses-noisedist-22}b, both scaled appropriately for each scale level in this plot as described in the text.}
\label{impulses-examplenoisedist}
\end{figure}

The results of this section can be used to asses the statistical significance of transform maxima.  This is illustrated in figure~\ref{impulses-examplenoisedist} for the example presented earlier in figure~\ref{impulses-morsetrain-noisy}.  Here we plot the scale locations $\hat s_n$ and magnitudes of all the transform maxima detected in the example, shown here with their non-normalized magnitudes $|\hat w_n|$ in (a) and normalized magnitudes $|\widetilde w_n|$ in (b); these are the gray dots together with the black circles.  For consistency with the spectra shown in figure~\ref{impulses-spectra}, we plot the effective period $2\pi/\omega_{\hat s_n}=2\pi \hat s_n/\omega_{\beta,\gamma}$ rather than the scale $\hat s_n$ itself.  The distributions and associated false detection rate appropriate for this length $M=12000$ time series are then determined by simulating 1000 $\times$ 12000 $\by_{\alpha,\beta,\gamma}(s)$-vectors for each scale $s$. The curves show the resulting expected detection rates for a time series of length $M=12000$.  The rates here are expressed as events per time series of length $M$, such that 1/1000 means that an event of the indicated magnitude or larger is expected at a particular scale only once per 1000$M$ or $1.2\times10^7$ data points.  Because there are 57 scale band being analyzed, events of a larger magnitude are expected to occur at {\em any} scale at a rate of one per $1000/57 M$ or about one per $20 M$.  

Choosing the 1/1000 rate as our cutoff, we find seven events exceeding this level of significance, corresponding to the maxima associated with the six events of the noise-free signal, plus one duplicate maxima associated with the second event.  This duplicate happens because the numerical algorithm has located two transform maxima very closely spaced together, a not uncommon occurrence.  From these seven statistically significant transform maxima, we estimate the properties of the underlying events using (\ref{inferences}), and reconstruct the signal by inserting these into the element model (\ref{morseelementmodel}).  The result, shown as the black dotted curve in figure~\ref{impulses-morsetrain-noisy}a, is very close to the original signal for most of the record and is therefore not visible.  However, in the vicinity of the second event, it overshoots the original signal on account of the duplicate maximum.  This difficulty is one of the reasons an isolation criterion is required, as developed in the next section.

In this subsection, an example of assessing statistical significance of events in a white noise background has been presented.  As another example, the case of $\alpha=1$ red noise is addressed in supplemental figures S1, S2, and S3, which are the red noise analogues of figures~\ref{impulses-morsetrain-noisy}, \ref{impulses-noisedist-22}, and \ref{impulses-examplenoisedist}, respectively.  See the captions of those figures for further discussion.

\subsection{Regions of influence}\label{subsection:region}

The final step is to introduce conditions for guaranteeing that the transform maxima are sufficiently isolated from one another, as well as from any regions of missing data.  There several reasons for doing so. Firstly, the element method depends upon the assumption that the events in the signal model (\ref{morseelementmodel}) are sufficiently isolated such that in the vicinity of transform maxima, the events may be regarded independently from one another.  In real-world applications, there may be sources of variability for which this is not the case, and the properties of such events are not expected to be accurately recoverable.  Therefore such events should be rejected from the event detection  results on account of being insufficiently isolated.  Secondly, discretization effects and/or noise can often lead to multiple closely-spaced transform maxima associated with the {\em same} event, which should not be taken to represent independent events.  In such cases, it is desirable to have a means of determining the primary transform maxima, and rejecting the others.  Finally, domain edges or missing data can also contribute to creating spurious maxima. 

In this section we will make use of an expansion of the time-domain wavelets as approximately consisting of a modulated Gaussian, 
\begin{equation}\label{cumulantexpansion}
\psi_{\beta,\gamma}(t)=
\psi_{\beta,\gamma}(0) \exp\left\{\ri t K_{1;\beta,\gamma} - \frac{1}{2} t^2K_{2;\beta,\gamma}\right\} +\epsilon_{3;\beta,\gamma}
\end{equation}
which has been examined in detail by \cite{lilly12b-itsp}.  The $K_{n;\beta,\gamma}$ quantities, termed the wavelet {\em cumulants}, are terms in a Taylor series of the natural logarithm of the wavelet, which here has been truncated after the second-order term.  The first- and second-order cumulants are given by
\begin{equation}\label{cumulantdefinitions}
K_{1;\beta,\gamma}\equiv\frac{\Gamma\left(\frac{\beta+2}{\gamma}\right)}{\Gamma\left(\frac{\beta+1}{\gamma}\right)},\quad\quad K_{2;\beta,\gamma} \equiv\frac{\Gamma\left(\frac{\beta+3}{\gamma}\right)}{\Gamma\left(\frac{\beta+1}{\gamma}\right)} - \left[\frac{\Gamma\left(\frac{\beta+2}{\gamma}\right)}{\Gamma\left(\frac{\beta+1}{\gamma}\right)}\right]^2
\end{equation}
see Appendix~B  and \S~III-A of \cite{lilly09-itsp}, or \S~S1 of the supplementary text.  $K_{1;\beta,\gamma}$ plays the role of a frequency, $1/\sqrt{K_{2;\beta,\gamma}}$ is the standard deviation of the Gaussian in the time domain, and $\epsilon_{3;\beta,\gamma}$ is an error term that is implicitly defined as a residual.  Because the wavelet magnitude $|\psi_{\beta,\gamma}(t)|$ has a roughly Gaussian profile, the second- expansion of the logarithm of the wavelet is a much better approximation than would be obtained by the second-order Taylor series of the wavelet itself.

A solution to determining whether the events are well isolated from one another is based on the expected {\em region of influence} associated with a transform maxima.  We will identify the curve at which the wavelet transform modulus has fallen off to some fraction $\lambda$ of its peak value.  Assuming an event with scale $\rho$ located at time $\tau=0$, we are interested in the ($\tau,s)$ curve satisfying 
\begin{equation}\label{simplephitildelevelsetdef}
\left|\zeta_{\beta,\mu,\gamma}(\tau/\rho,s/\rho) \right|=\left|\zeta_{\beta,\mu,\gamma}(\tilde\tau,\tilde s) \right|=\lambda \zeta_{\beta,\mu,\gamma}^{\max}
\end{equation}
where again $\tilde \tau =\tau/\rho$ and $\tilde s = s/\rho$. Knowledge of the Morse wavelets allow us to readily obtain a closed-form expression for an approximation to this curve.  Inserting the cumulant expansion (\ref{cumulantexpansion}) into the expression for $\zeta_{\beta,\mu,\gamma}(\tau,s,\rho)$ as a wavelet given earlier in (\ref{simplephi}), (\ref{simplephitildelevelsetdef}) becomes
\begin{equation}\label{simplephitildelevelsetexpansion}
\frac{\tilde s^{\beta}}{\left(\tilde s^\gamma+1\right)^{(\beta+\mu+1)/\gamma}}\exp\left\{ - \frac{1}{2} \left(\frac{\tilde \tau}{\sqrt[\gamma]{\tilde s^\gamma+1}}\right)^2K_{2;\beta+\mu,\gamma}\right\}\approx\lambda \vartheta_{\beta,\mu,\gamma}
\end{equation}
after also making use of (\ref{zetamaxdef}).  Here we have chosen to ignore the error term $\epsilon_{3;\beta,\gamma}$ in the cumulant expansion arising from terms higher than second order.  This rearranges to give
\begin{equation}\label{simplephitildelevelset}
\tilde \tau \approx \pm \left[2\frac{(\tilde s^\gamma+1)^{2/\gamma}}{K_{2;\beta+\mu,\gamma}} \ln
\left( \frac{\tilde s^{\beta}}{\lambda \vartheta_{\beta,\mu,\gamma}\left(\tilde s^\gamma+1\right)^{(\beta+\mu+1)/\gamma}}\right)\right]^{1/2}
\end{equation}
as an approximation to the region of influence for a $(\mu,\gamma)$ Morse function analyzed with a $(\beta,\gamma)$ wavelet, and using the $1/s$ scale normalization in the wavelet transform.

This region of influence expression can readily be evaluated numerically.  The right-hand-side of (\ref{simplephitildelevelset}) is real-valued for the region of scales over which the numerator in the natural logarithm exceeds the denominator. While the exact locations of the crossover points of these two curves do not have convenient analytic expressions, analyzing their behaviors shows that the range of scales for which (\ref{simplephitildelevelset}) is real-valued occurs within the somewhat broader range
\begin{equation}\label{endpointscales}
 \left(\lambda \vartheta_{\beta,\mu,\gamma}\right)^{1/\beta}<\tilde s< \left(\frac{1}{\lambda \vartheta_{\beta,\mu,\gamma}}\right)^{1/\left(\mu+1\right)}
\end{equation}
as shown in Appendix~D.  Therefore, to compute the $\tilde \tau$ curves, we determine the two endpoint scales in (\ref{endpointscales}), form an array of normalized scales $\tilde s$ over this range, compute (\ref{simplephitildelevelset}), and then omit any end regions in which $\tilde \tau$ is found to take on imaginary values.

The regions of influence are employed in the element analysis as follows. After identifying a set of transform maxima, and excluding those falling below a certain significance level based on the noise model, we then exclude those that are not sufficiently isolated.  To do that, we choose a certain $\lambda$ level, for example $\lambda=1/2$, and compute the approximate regions of influence for each transform maximum by appropriately shifting and rescaling (\ref{simplephitildelevelset}). The transform maxima are sorted in order of decreasing amplitude, and a maxima point is rejected if any larger-amplitude maxima points are found to occur within its own region of influence, as this would indicate that it is not well isolated.  The remaining maxima are said to be isolated at the particular $\lambda$ level.

Finally, to deal with edge effects and the influence of missing data, the following approach is adopted.  All gaps are first linearly interpolated over, and locations of missing data are recorded.  When transform maxima are detected, the fraction of missing data, or data points outside the time series boundaries, is determined over a time period one wavelet footprint in duration centered on each maxima.  Transform maxima containing more than some percentage, say 10\%, missing data are then rejected.  This approach allows missing data segments of any length to be dealt with, while at the same time utilizing as much information as possible from the data.  The missing data condition should be applied before the isolation criterion, in order to prevent spurious maxima arising from missing data effects to interfere with physically meaningful maxima.

In the example of figure~\ref{impulses-morsetrain-noisy}, as described earlier, seven statistically significant maxima are detected.  However, computing the regions of influence using (\ref{simplephitildelevelset}) based on the known transform maxima location $(\hat\tau_n,\hat s_n)$ together with $\beta$, $\mu$, and $\gamma$, we find one of the two maxima in the vicinity of the second event is not well isolated at the $\lambda=1/2$ level.  Rejecting this event, we are left with the six events shown as black circles in figures~\ref{impulses-morsetrain-noisy} and \ref{impulses-examplenoisedist}.  In the former figure, the $\lambda=1/2$ regions of influence around the six significant and isolated transform maxima are shown.  In real-world applications, this isolation criterion is found to be crucial for obtaining good performance.

Limiting the reconstruction using (\ref{morseelementmodel}) to these six points, we obtain the red curve shown in figure~\ref{impulses-morsetrain-noisy}.  Despite the very noisy appearance of the analyzed signal in figure~\ref{impulses-morsetrain-noisy}, the original events are detected with a very high degree of statistical significance, and the reconstruction is virtually identical to the original signal.    This illustrates that the element analysis can accurately extract signals of the form (\ref{morseelementmodel}) even in the presence of relatively large noise levels.  

A natural question is the extent to which events may be obscured by other nearby events.  This is explored in supplementary figure~S4 for the noise-free version of the signals shown in figure~\ref{impulses-morsetrain-noisy}, plus a set of closely-spaced smaller-magnitude events.  It is seen that the smaller-amplitude events may be detected provided they are not too close to the large-amplitude events, with the region of influence of the larger-amplitude events providing some guidance as to the shielding region.  A more complete investigation of such effects is beyond the scope of this paper. 


\section{Application}\label{section:application}


An application to real-world data is presented in figure~\ref{impulses-labseathreepanel}.  The dataset analyzed here is a small segment of along-track data from the TOPEX/Poseidon/Jason/Ocean Surface Topography Mission satellite altimeters, and consists of 5216 valid data points.  We use a reprocessed, homogeneous along-track dataset, the Integrated Multi-Mission Ocean Altimeter Data for Climate Research, Version 3 dataset \cite{beckley10-mg,beckley}.  The ground tracks in this dataset are repeated exactly every 9.92 days, and have an along-track resolution of about $\Delta=5.7$~km at the latitude considered here.  The quantity measured is sea level anomaly relative to an unknown temporal mean.  

One year's worth of data is shown from a particular track in the Labrador Sea, a small marginal sea located between Greenland and Canada.  The Labrador Sea is a well-known area of energetic coherent eddies \cite{eden02-jpo,gelderloos11-jpo,luo11-pio,dejong14-jpo},  which were the subject of a study using along-track data in an early {\em ad hoc} prototype of the method developed here \cite{lilly03-pio}.   The particular track chosen crosses the Labrador Sea from southwest to northeast, passing within about 12~km of the site of the historical ``Bravo'' current meter mooring, see figure~24 of \cite{lilly03-pio}.  Southwestern locations are at the left, and northeastern locations are at the right.  The gap in the lower left of this figure is due to the seasonal advance of sea ice from the coastal Labrador Current, which interferes with the altimetric measurements.  The upwards bumps seen in the central part of the track are the signatures of long-lived coherent eddies, and are the structure we wish to detect and quantify.  



In keeping with the use of a Gaussian as a model for eddies, as is standard in the literature, the element function used for this dataset is the analytic Gaussian $\psi_{0,2}(t)$.  The time series are analyzed using a $\psi_{1,2}(t)$ wavelet, with additional parameter settings as given in Appendix~C. The noise is taken to be Gaussian white noise.  From the variance within the highest frequency band, which corresponds to a period $2\pi/\omega_{s_1}$ of 5.5 data points, we infer from (\ref{sigmaeps}) a noise standard deviation of $\sigma_\epsilon=$3.2~cm.  This is used to assess statistical significance levels using the simulation method described in \S~\ref{subsection:region}.  A very high level of significance is chosen, such that in each frequency band, events with a false detection rate greater than one event per 1000 realizations of this dataset are rejected.  Using the region of influence condition, maxima are rejected if they are not isolated at the $\lambda=1/2$ level, and also if they contain more than 10\% missing data.

The above steps lead to a small number of detected events, 67 altogether or less than two per track, that are determined to be highly statistically significant as well as isolated from one another and from missing data segments. Reconstructions based on the element analysis method are shown in the central panel.  These are seen to explain virtually all of the meaningful structure. A persistent eddy feature of about 20~km in radius is clearly observed in the upper half of the central panel. The residuals (originals minus reconstructions) are shown at the right, and appear virtually devoid of meaningful structure, showing that the model is indeed a good fit to the observations.  Moreover, as the detected events reconstruct the data using only  5\% ($4\times 67 / 5216\approx 0.05)$ as many coefficients as there are datapoints, the information within the data is represented with a high degree of compression.

In an earlier study using  a prototype version of this method, the detected events in alongtrack altimetry were analyzed in detail to understand the physical properties of coherent eddies in the Labrador Sea, see \S~5--6 of  \cite{lilly03-pio}.  In particular, validation against {\em in situ} velocity measurements from a moored current meter was carried out in order to ensure that the detected events were physically meaningful, see figure~33 in that paper.  As the events here appear qualitatively similar to the previously identified events, they are also likely to be physically meaningful.  The earlier study used data from 1992 until about mid-2000, thus an extension of that study using a longer data record would be valuable in order to assess interannual variability in eddy statistics.  However, this would require a considerable amount of more work and is outside the scope of the present paper, which is limited to the development of the method.  Further analysis of the detected events is left to a sequel.  The main new result is the ability to detect events using a method that can rigorously assess event significance, which was done in an {\em ad hoc} basis in the earlier study.  

This application shows that coherent eddy properties as small as $\mathcal{O}$(10)~km can indeed be extracted from the along-track dataset. This data segment analyzed here represents only one-thousandth of one percent of the over 400 million data points within the entire along-track altimeter dataset.  Considering the vast size of the complete dataset underscores the need for a fully automated method, and justifies the effort that has been required in its development.  The element method makes possible a global eddy census along the lines of \cite{chelton11-pio}, but with the ability  to resolve features an order of magnitude smaller  than has previously been possible.


\section{Conclusions and Discussion}

A method has been developed for analyzing time series that consist of rescaled, phase-shifted, isolated replicates of a specified time-localized function.  Time series of this type could be described as being ``impulsive'' in nature, as opposed to singular or oscillatory.  The method, termed element analysis, is inspired by the continuous wavelet transform, and utilizes the generalized Morse wavelet family as both a basis and an analysis tool.  The element model is intended as a third major category of wavelet-based signal model, complementing the wavelet ridge and modulus maxima methods by allowing signals to be supported only at isolated points on the time/scale plane.  Particular innovations are the creation of a simplified framework for efficient simulation of maxima statistics, as well as the identification of an approximate form for the regions of influence on the time/scale plane.  While the method was formulated for real-valued time series, the extension to complex-valued time series is straightforward.

The  method is applied to the detection of coherent eddy events in along-track satellite altimetry, with encouraging results.  Furthermore, there is good reason to believe that this method may be useful in a wide range of problems.  In addition to being suitable for eddy detection, the method is also appropriate for strongly modulated wave packets or other impulse-like events, common features in many physical systems.  Moreover, the method can be seen in some respects as a generalization of a Fourier series representation, with the beneficial aspect of allowing for time localization in signal features.  In comparison with the wavelet thresholding method, element analysis is more specific because it requires the support points to be isolated from one another, thus filtering out events for which the proposed signal form is not a good match.

There are a number of obvious ways that the method presented here could be extended.  Firstly, while the shape of the normalized detection rate was found to collapse with a suitable normalization, as shown in figure~\ref{impulses-noisedist-22}b, the total detection rate---the $y$-intercept of these curves---takes on a range of values.  When examined over the $\beta$ and $\gamma$ plane (not shown), there emerges what appears to a meaningful pattern in the total detection rate, but the reasons for this variation is not clear.  Thus better understanding the distribution of noise events from first principles is a direction for future work.  Secondly, it would be straightforward to work out expressions for the bias and variance associated with the estimated event properties in the presence of noise, if desired.  Thirdly, there is the question of how to choose an element wavelet, i.e. determining the values of $\mu$ and $\gamma$ that best capture signal structure, and the sensitivity of the analysis to this choice.  Finally, the model proposed here, while fairly flexible, could be generalized still further by allowing each event to be composed of a superposition of the higher-order orthogonal versions of the Morse wavelets that emerge from the localization region formalism \cite{daubechies88-ip,olhede02-itsp,olhede03a-prsla}.  This may involve augmenting the element method with the polarization analysis of \cite{lilly95-gji,olhede03a-prsla}.  Alternatively, these higher-order wavelets could be used as the basis for a more refined metric for determining the local quality of a fit, by excluding events which project too strongly onto the next few orthogonal wavelets in the vicinity of a transform maximum.  This could represent an additional means of classifying the properties of the detected events, which would aid in their interpretation.

\appendix

\section*{Appendix A. A freely available software package}\label{appendix:jlab}

This appendix presents some notes on a numerical implementation of the element method distributed in the Matlab toolbox \texttt{jLab}, available at \url{http://www.jmlilly.net}.  In \texttt{jLab}, the Morse wavelet frequency array is determined by \texttt{morsespace} using the criteria discussed in Appendix~C.  The wavelet transform is then implemented by \texttt{wavetrans}.  The wavelets themselves are computed by \texttt{morsewave}, which is called internally by \texttt{wavetrans}.  In the discrete implementation, the maxima conditions (\ref{maxconditions}) may be replaced with criteria to locate time/scale points larger in magnitude than the four neighboring points in time and in scale.  In \texttt{jLab}, this is done with the \texttt{transmax} routine, leading to an array of indices into maxima times $\hat\tau$ and scales $\hat s$, transform values $\hat w$ that are quadratically interpolated between discrete scale levels, and scale frequency values $\omega_{\hat s}$ corresponding to the maxima locations that are similarly interpolated.   \texttt{transmax} also identifies the missing data fraction over one wavelet footprint centered on each maxima point. The conversion of the transform maxima values to the element function parameters as in (\ref{inferences}) is carried out by the routine \texttt{maxprops}. 

Simulations of the histograms of transform maxima due entirely to noise, and the associated false detection rates, are performed by \texttt{transmaxdist} using 5-vector method described in \S~\ref{section:noisesim}.  Both the regions of influence and the localization regions are computed by \texttt{morseregion}, based on the analytic expressions given in (\ref{simplephitildelevelset}) and in \cite{olhede03a-prsla} respectively.  The routine \texttt{isomax} is then used to verify that the maxima are isolated from one another.  This routine calls \texttt{morseregion} to compute the regions of influence around a set of maxima as output by \texttt{transmax}, and returns a boolean array that is true for those points that do not encompass a larger-magnitude maxima within their localization regions for a specified $\lambda$ value. The region of influence curves associated with each maxima are also output by \texttt{isomax}. 

Finally, all analysis and figure generation associated with this paper are carried out by the script \texttt{makefigs\_element}.  Running this script takes about nine minutes on a 12-core Mac Pro with 2.7 GHz Intel Xeon E5 processors.  Most of this time is due to the large Monte Carlo simulations that are carried out in investigating the noise distributions; the application to data, which includes running the analysis method on a dataset excerpt about 20 times the size of that shown in figure~\ref{impulses-labseathreepanel}, takes about two minutes.  




\section*{Appendix B. The wavelet footprint vs. standard deviation}

In this appendix, the relationship between the wavelet's footprint $L_{\beta,\gamma}(s)$, defined in (\ref{Ldef}), and its time-domain standard deviation is discussed.  A conventional measure of the wavelet duration, related to $P_{\beta,\gamma}$, is the time-domain standard deviation 
\begin{equation}
\sigma_{t;\beta,\gamma}^2 \equiv \omega_{\beta,\gamma}^2 \frac{\int_{-\infty}^{\infty} t^ 2|\psi_{\beta,\gamma}(t)|^ 2\,\rd t}{\int_{-\infty}^{\infty}|\psi_{\beta,\gamma}(t)|^ 2\,\rd t}\label{timespread}
\end{equation}
which is defined here to be dimensionless.  The dimensional time-domain standard deviation of the scale $s$ wavelet is therefore given by  $s\sigma_{t;\beta,\gamma}/\omega_{\beta,\gamma}$.  Whereas the value of $P_{\beta,\gamma}$ has the simple expression $P_{\beta,\gamma}=\sqrt{\beta\gamma}$, the analytic expression for $\sigma_{t;\beta,\gamma}$ for the generalized Morse wavelets is somewhat complicated, see eqn.~(47) of \cite{lilly09-itsp}.  However, numerical calculations show $P_{\beta,\gamma}$ is roughly equal to $\sqrt{2}\sigma_{t;\beta,\gamma}$ over a large range of $\beta$ and $\gamma$ values.  The factor of $2\sqrt{2}$ in the definition  (\ref{Ldef}) of the wavelet footprint $L_{\beta,\gamma}(s)$ thus makes $L_{\beta,\gamma}(s)$ comparable to four times the scale $s$ wavelet's dimensional time-domain standard deviation, $s\sigma_{t;\beta,\gamma}/\omega_{\beta,\gamma}$.

This link between $P_{\beta,\gamma}$ and $\sigma_{t;\beta,\gamma}$ further justifies the interpretation of $P_{\beta,\gamma}$ as an inverse bandwidth measure, mentioned in the text after (\ref{pdef}), because the Heisenberg uncertainty principle bounds the product of the time-domain and frequency-domain standard deviations.\footnote{The author is grateful to Wayne King for this insight. } 

\section*{Appendix~C: Wavelet transform implementation details}\label{section:implementation}

In the numerical implementation employed herein, wavelets are computed by specifying $\beta$ and $\gamma$ together with their scale frequency $\omega_s\equiv\omega_{\beta,\gamma}/s$, as opposed to their scale $s$.  Furthermore, the convention is adopted that the sampling interval $\Delta$ is always interpreted as being unity when referring to the scale frequencies. There are several details regarding the choice of an array of scale frequencies for the transform that are relevant to mention.  The first pertains to the choice of high-frequency cutoff.  Choosing some positive number $\eta<1$, we determine the highest acceptable scale frequency $\omega_s=\omega_{\mathrm{high}}$ to be used in the transform as the smallest choice of $\omega_{\mathrm{high}}$ such that
\begin{equation}\label{omegahigh}
\Psi_{\beta,\gamma}\left(\pi\frac{\omega_{\beta,\gamma}}{\omega_{\mathrm{high}}}\right)\leq \eta \,\Psi_{\beta,\gamma}(\omega_{\beta,\gamma})
\end{equation}
which indicates that, for a wavelet characterized by scale $s=\omega_{\beta,\gamma}/\omega_{\mathrm{high}}$, the value of the wavelet at the Nyquist frequency $\omega=\pi$ will have decayed to a value no greater than $\eta$ times its peak value. Wavelets with scale frequencies $\omega_s>\omega_{\mathrm{high}}$ will extend substantially past the Nyquist.

In wavelet analysis it is standard to designate a frequency array such that its logarithm is  linearly spaced.  It is desirable to choose the frequency spacing in such a way that the frequency resolution is compatible with the bandwidth.   Numerical problems can arise if the frequency array is either too coarsely spaced or too finely spaced relative to the bandwidth.  We set the $j$th scale frequency in terms of the highest frequency $\omega_{\mathrm{high}}$ and a density parameter $D$ as
\begin{equation}\label{frequencyarray}
\omega_{s_j} = \frac{\omega_{\mathrm{high}}}{\left(1+\frac{1}{D P_{\beta,\gamma}}\right)^{j-1}},\quad\quad\quad  \frac{\omega_{s_{j-1}}-\omega_{s_{j}}}{\omega_{s_{j}}} =\frac{1}{D }\frac{1}{P_{\beta,\gamma}}
\end{equation}
which orders the frequencies in decreasing order, with $\omega_{s_1}=\omega_{\mathrm{high}}$.  The second expression gives the fractional difference between two successive scale frequencies, such that $D$ can be seen as the number of distinct scale frequency bands that fit within a frequency interval of width $\omega_{s_{j}}/P_{\beta,\gamma}$ around the $j$th wavelet.  We find $D=4$ or $D=8$ to generally be suitable choices.

In vector-based noise transform simulations of \S~\ref{section:noisesim}, especially in (\ref{covdef})--(\ref{bigSigma}), a factor of $r$ appears which is the ratio of successive scales.  We let this take on the value implied by (\ref{frequencyarray})
\begin{equation}
r=\frac{s_{j+1}}{s_j} = 1+\frac{1}{DP_{\beta,\gamma}}
\end{equation} 
in agreement with the scale discretization scheme for the wavelet transform.

To determine the lowest scale frequency  $\omega_s=\omega_{\mathrm{low}}$, we specify that some number $p$ wavelets, as measured by the wavelet footprint  $L_{\beta,\gamma}(s)$,  should fit within a length $M$ time series. That is we set $p L_{\beta,\gamma}(s) =M$, which from (\ref{Ldef}) determines the lowest scale frequency to be
\begin{equation}\label{omegalow}
\omega_{\mathrm{low}}=p2\sqrt{2} P_{\beta,\gamma}/M.
\end{equation} 
The number of frequency bands $J$ can be found by using (\ref{frequencyarray}) to determine the largest integer $J$ such that $\omega_{s_J}$ is no smaller than $\omega_{\mathrm{low}}$.  A typical choice for $p$, called the ``packing number'', is $p=5$, specifying that in the lowest frequency band, five wavelet footprints span the time series.

At the edges of the time series, some boundary condition must be applied in order for the transform to be well-defined.  Typical choices are setting the time series to zero outside of its boundaries, a periodic boundary condition, or a ``mirror'' condition in which the time series is extended by flipping it about its two endpoints.  The latter generally performs performs much better than the others in terms of minimizing edge effects, and so this is what will be used here.

The wavelet transform for the synthetic example presented in figure~\ref{impulses-morsetrain-noisy} is taken at 59 logarithmically-spaced scale levels, with a cutoff parameter $\eta=0.05$ from (\ref{omegahigh}) setting the highest frequency, an overlap factor $D=4$ in (\ref{frequencyarray}) determining the frequency spacing, and a lowest frequency specified by a packing number $p=3$ in (\ref{omegalow}). For the application in \S~\ref{section:application}, the wavelet transform is taken in 38 frequency bands, with parameter choices $\eta=0.1$, $D=8$, and $p=2$.

\section*{Appendix D. A scale range for the region of influence}

In this appendix we determine a range of scales bounding the range over which the region of influence curves (\ref{simplephitildelevelset}) have a real-valued solution.  The range of scales that needs to be computed can be determined as follows.  For convenience here, we introduce the notation 
\begin{equation}f_1(\tilde s)\equiv \tilde s^{\beta},\quad\quad f_2(\tilde s) \equiv\lambda \vartheta_{\beta,\mu,\gamma} \left(\tilde s^\gamma+1\right)^{(\beta+\mu+1)/\gamma}.\end{equation}
Note that both curves increase monotonically with $\tilde s$.  For small $\tilde s$, $f_1(\tilde s)$ vanishes while $f_2(\tilde s)$  tends to $\lambda \vartheta_{\beta,\mu,\gamma}$ from above, thus $f_1(\tilde s)<f_2(\tilde s)$ for small $\tilde s$.  Because  $f_2(\tilde s) > \lambda \vartheta_{\beta,\mu,\gamma}$, the smallest value of  $\tilde s$ at which $f_1(\tilde s)$ rises above $f_2(\tilde s)$ will be no smaller than $\tilde s_a \equiv (\lambda \vartheta_{\beta,\mu,\gamma})^{1/\beta}$, the location at which $f_1(\tilde s)$ equals $\lambda \vartheta_{\beta,\mu,\gamma}$.  For large $\tilde s$, we have $f_2(\tilde s)\sim \tilde s^{\beta+\mu+1}$ which grows faster with $\tilde s$ than $f_1(\tilde s)= \tilde s^{\beta}$.  Thus for sufficiently large $\tilde s$ we must again have $f_1(\tilde s)<f_2(\tilde s)$.  From the inequality  $f_2(\tilde s) > \lambda \vartheta_{\beta,\mu,\gamma} \tilde s^{\beta+\mu+1}$, the smallest value of $\tilde s$ at which $f_1(\tilde s)$ again falls below $f_2(\tilde s)$ will be no larger than $\tilde s_b \equiv \left[1/(\lambda \vartheta_{\beta,\mu,\gamma})\right]^{1/\left(\mu+1\right)}$.  Thus the scales for which $f_1(\tilde s)$ exceeds $f_2(\tilde s)$ fall between a smallest scale of  $\tilde s_a \equiv (\lambda \vartheta_{\beta,\mu,\gamma})^{1/\beta}$ and a largest scale of  $\tilde s_b = \left[1/(\lambda \vartheta_{\beta,\mu,\gamma})\right]^{1/\left(\mu+1\right)}$, as claimed.

\section*{Appendix~E.  Regions of influence vs. localization regions} 

\begin{figure}[t!]
\begin{center}
\includegraphics[height=17.5cm,angle=-90]{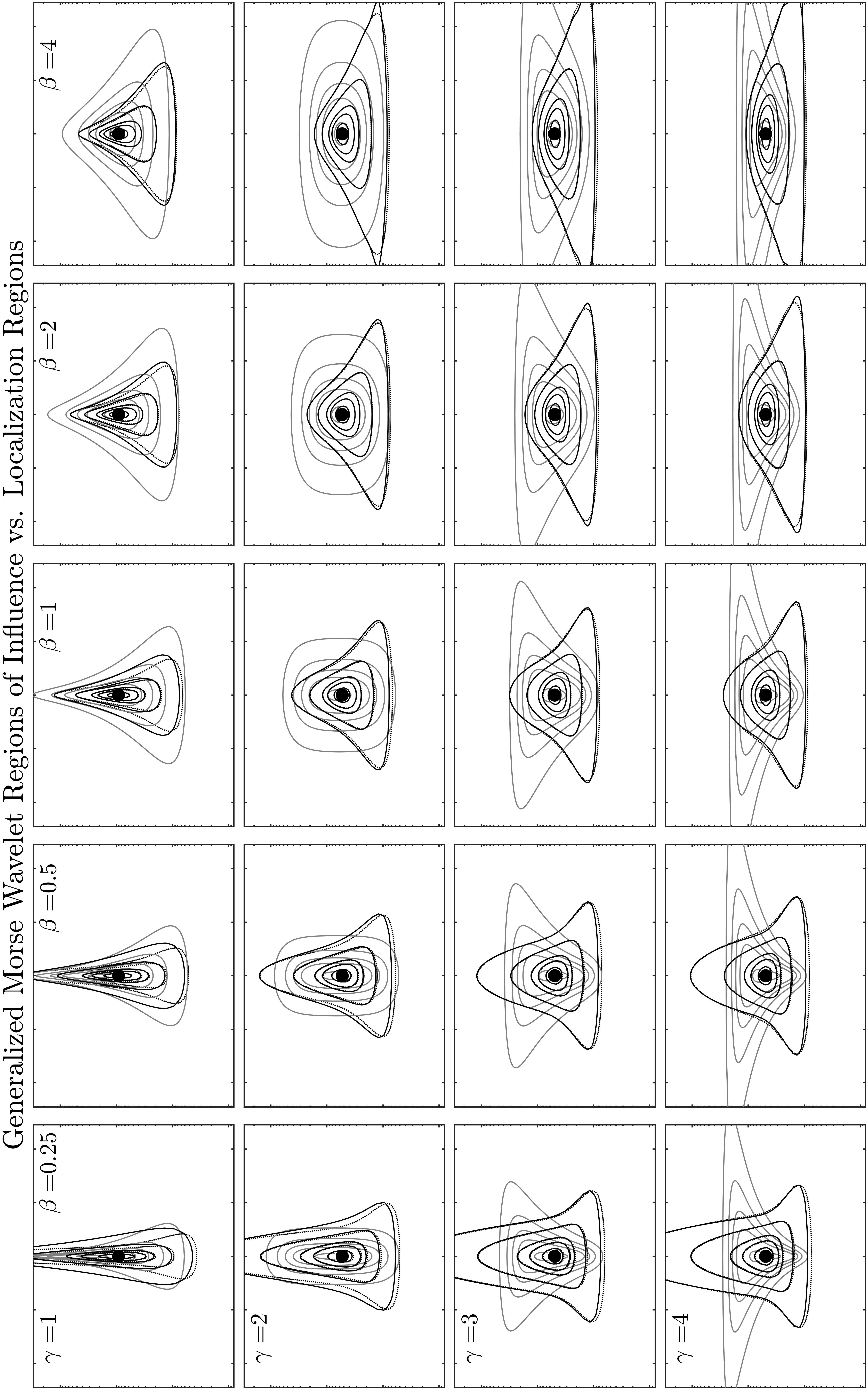}\end{center}
\caption{Regions of influence for the wavelet transform of a $(0,\gamma)$ Morse function, transformed by Morse wavelets in the same $\gamma$ family, compared with the Morse wavelet localization regions.    Here, rows correspond to four different $\gamma$ values, and columns to five different $\beta$ values of the analyzing wavelet, as labeled; this is just as in figure~\ref{timedomainwavelets}. The element function is $\psi_{0,\gamma}(t/\rho)$, with $\rho$ chosen such that the element function scale frequency is $\omega_{0,\gamma}/\rho = 2\pi/100$.  Regions of influence for five different $\lambda$-values are shown: $\lambda=0.25$, 0.5, 0.75, 0.85, and 0.95, with the higher values being closer to the center of each plot.  Solid black lines are the actual $\lambda$-contours from the numerically computed wavelet transform, while dotted black lines show the approximations based on the second-order cumulant expansion given in (\ref{simplephitildelevelset}); these are generally visually identical apart from the outermost curve at $\lambda=0.25$.  Thin gray lines show the localization regions, described in the text, of a wavelet having scale frequency $\omega_s= 2\pi/100$, with the area parameter $A_{\beta,\gamma}$ taking on the values  of 1, 10, 20, 40, and 100.  The axis limits are the same for each subplot. The $x$-tick marks are at times $\tau=-200$, -100 0, 100, and 200, while the major $y$-tick marks are at scale frequencies $\omega_s=1/100$, 1/10, and 1. Note that the $y$-axes are logarithmic.}
\label{regionsofinfluence}
\end{figure}

This appendix examines the Morse wavelet regions of influence in more detail, and compares them with another type of region, the localization regions of \cite{daubechies88-ip}.  Figure~\ref{regionsofinfluence} shows regions of influence with the element functions set to be the zeroth-order Morse functions $\psi_{0,\gamma}(t)$ for four different choices of $\gamma$, presented earlier in the left-hand column of figure~\ref{timedomainwavelets}.  The wavelet transforms of these functions are taken with each of five different analyzing wavelets in the same $\gamma$ family, corresponding in figure~\ref{timedomainwavelets} to all the other wavelets in the same row. Five $\lambda$-levels ($\lambda=0.25$, 0.5, 0.75, 0.85, and 0.95) are computed from the wavelet transforms, and are shown along with the approximate $\lambda$-levels from (\ref{simplephitildelevelset}).  Apart from the outermost curve for $\lambda=0.25$, the numerically computed contour and the approximation are visually indistinguishable.  The relatively poor behavior for small $\lambda$ values is not unexpected, due to the omission of higher-order terms in the expansion.  Thus provided $\lambda$ is not too small, the $\lambda$-levels are well approximated by (\ref{simplephitildelevelset}).

The general shape of the curves is similar in each case to an upward-pointing wedge or arrowhead, which becomes more oval in shape as $\lambda$ increases and the contours decrease in size. Moving left to right, as the order $\beta$ of the analyzing wavelet increases with a fixed element function, the aspect ratio of the transform $\lambda$-curves shifts from being more elongated in frequency to being more elongated in time. Moving from top to bottom, as the element function changes with the analyzing wavelet held fixed, the transforms are more similar to one another. A change in behavior is seen across the frequency at which the transform obtains its maximum modulus. Particularly for low $\beta$ values, the transform is seen to expand outward for lower frequencies compared to higher frequencies, consistent with the asymptotic behavior examined in (\ref{simplephilimits}).

Another type of region of the time/frequency plane that may be associated with the Morse wavelets is the {\em Morse wavelet localization region}. The original derivation of the Morse wavelets, in Appendix~A  of \cite{daubechies88-ip}, constructed them as solutions to a time/frequency localization problem; see also \cite{bayram00-nnsp,olhede02-itsp,olhede03a-prsla}.  Specifically, when the wavelets are reconstructed from their own transforms with themselves, limiting the reconstruction integral to some region leads to an eigenvalue problem, in which the eigenvalue can be thought of as an energy  concentration fraction within the region.  The Morse wavelets are the solutions to this eigenvalue problem when a particular choice---dependent upon $\beta$ and $\gamma$, as well as an area parameter  $A_{\beta,\gamma}$---is made for the localization region.  Convenient expressions for the Morse wavelet localization region may be found in \S~2(a) of \cite{olhede03a-prsla}.  The version of $A_{\beta,\gamma}$ used here is one-sided and therefore differs from that of \cite{olhede03a-prsla} by a factor of one-half. 

For comparison with the regions of influence, figure~\ref{regionsofinfluence} also shows the Morse wavelet localization regions, for five choice of area ($A_{\beta,\gamma}=1$, 10, 20, 40, and 100).  The localization regions are strongly dependent upon $\gamma$: for $\gamma<2$ the localization regions point upwards, as do the regions of influence, but for $\gamma>2$ they flip and point downwards.   For $\gamma=2$ they are nearly symmetric about their centers along this logarithmic frequency axis.  Thus, the localization regions are not at all appropriate as indicators of the regions of influence.  These two types of regions are different in principle.  The localization regions are about the {\em reconstruction integral}, not the transform, whereas the regions of influence indicate the amplitude of the transform on the time/frequency plane, and therefore depend upon the choice of scale normalization.  Furthermore, the localization regions are entirely a property of the analyzing wavelet, whereas the regions of influence depend upon the analyzing wavelet as well as the analyzed signal or element function.

\section*{Acknowledgments}

This work was supported by NASA contract \#NNH14CL78C. The author gratefully acknowledges helpful and inspiring interactions with Sofia Olhede, Jun Tong, Alfred Hanssen, P\aa l Erik Isachsen, and Wayne King in various stages of this work, as well as helpful comments from two anonymous reviewers. The Integrated Multi-Mission Ocean Altimeter Data for Climate Research Version 3 data were obtained from the NASA EOSDIS Physical Oceanography Distributed Active Archive Center (PO.DAAC) at the Jet Propulsion Laboratory, Pasadena, CA (\url{http://podaac.jpl.nasa.gov/dataset/MERGED_TP_J1_OSTM_OST_CYCLES_V3}). 
   


\begin{thebibliography}{35}
\providecommand{\natexlab}[1]{#1}
\providecommand{\url}[1]{\texttt{#1}}
\expandafter\ifx\csname urlstyle\endcsname\relax
  \providecommand{\doi}[1]{doi: #1}\else
  \providecommand{\doi}{doi: \begingroup \urlstyle{rm}\Url}\fi

\bibitem[Barthlott et~al.(2007)Barthlott, Drobinski, Fesquet, Dubos, and
  Pietras]{barthlott07-blm}
C.~Barthlott, P.~Drobinski, C.~Fesquet, T.~Dubos, and C.~Pietras.
\newblock Long-term study of coherent structures in the atmospheric surface
  layer.
\newblock \emph{Bound.-Lay. Meteorol.}, 125\penalty0 (1):\penalty0 1--24, 2007.

\bibitem[Bayram and Baraniuk(2000)]{bayram00-nnsp}
M.~Bayram and R.~Baraniuk.
\newblock \emph{Nonlinear and Nonstationary Signal Processing}, chapter
  Multiple window time-varying spectrum estimation, pages 292--316.
\newblock Cambridge University Press, 2000.

\bibitem[Beckley et~al.(2016)Beckley, Zelensky, Holmes, Lemoine, Ray, Mitchum,
  Desai, and Brown]{beckley}
B.~Beckley, N.~Zelensky, S.~Holmes, F.~Lemoine, R.~Ray, G.~Mitchum, S.~Desai,
  and S.~Brown.
\newblock Integrated multi-mission ocean altimeter data for climate research
  version 3.1.
\newblock Dataset accessed 2016-09-26., 2016.
\newblock URL
  \url{http://podaac.jpl.nasa.gov/dataset/MERGED_TP_J1_OSTM_OST_CYCLES_V3}.

\bibitem[Beckley et~al.(2010)Beckley, Zelensky, Holmes, Lemoine, Ray, Mitchum,
  Desai, and Brown]{beckley10-mg}
B.~D. Beckley, N.~Zelensky, S.~Holmes, F.~Lemoine, R.~Ray, G.~Mitchum,
  S.~Desai, and S.~Brown.
\newblock Assessment of the {J}ason-2 extension to the {TOPEX}/{P}oseidon,
  {J}ason-1 sea-surface height time series for global mean sea level
  monitoring.
\newblock \emph{Mar. Geod.}, 33\penalty0 ({S}1):\penalty0 447--471, 2010.
\newblock {S}upplemental {I}ssue on {OSTM}/{J}ason-2 calibration/validation.

\bibitem[Byrne et~al.(1995)Byrne, Gordon, and Haxby]{byrne95-jpo}
D.~A. Byrne, A.~L. Gordon, and W.~F. Haxby.
\newblock Algulhas eddies: a synoptic view using {G}eosat {ERM} data.
\newblock \emph{J. Phys. Oceanogr.}, 25:\penalty0 902--917, 1995.

\bibitem[Chelton et~al.(2011)Chelton, Schlax, and Samelson]{chelton11-pio}
D.~B. Chelton, M.~G. Schlax, and R.~M. Samelson.
\newblock Global observations of nonlinear mesoscale eddies.
\newblock \emph{Prog. Oceanogr.}, 91\penalty0 (2):\penalty0 167--216, 2011.

\bibitem[Chen et~al.(2001)Chen, Donoho, and Saunders]{chen01-siam}
S.~S. Chen, D.~L. Donoho, and M.~A. Saunders.
\newblock Atomic decomposition by basis pursuit.
\newblock \emph{SIAM Rev.}, 43\penalty0 (1):\penalty0 129--159, 2001.

\bibitem[Daubechies and Paul(1988)]{daubechies88-ip}
I.~Daubechies and T.~Paul.
\newblock Time-frequency localisation operators: a geometric phase space
  approach {II}. {T}he use of dilations and translations.
\newblock \emph{Inverse Probl.}, 4:\penalty0 661--80, 1988.

\bibitem[de~Jong et~al.(2014)de~Jong, Bower, and Furey]{dejong14-jpo}
M.~F. de~Jong, A.~S. Bower, and H.~H. Furey.
\newblock Two years of observations of warm-core anticyclones in the {L}abrador
  {S}ea and their seasonal cycle in heat and salt stratification.
\newblock \emph{J. Phys. Oceanogr.}, 44\penalty0 (2):\penalty0 427--444, 2014.

\bibitem[Delprat et~al.(1992)Delprat, Escudi\'e, Guillemain, Kronland-Martinet,
  Tchamitchian, and Torr\'esani]{delprat92-itit}
N.~Delprat, B.~Escudi\'e, P.~Guillemain, R.~Kronland-Martinet, P.~Tchamitchian,
  and B.~Torr\'esani.
\newblock Asymptotic wavelet and {G}abor analysis: {E}xtraction of
  instantaneous frequencies.
\newblock \emph{IEEE T. Inform. Theory}, 38\penalty0 (2):\penalty0 644--665,
  1992.

\bibitem[Donoho(1995)]{donoho95-itit}
D.~L. Donoho.
\newblock De-noising by soft-thresholding.
\newblock \emph{IEEE T. Inform. Theory}, 41\penalty0 (3):\penalty0 613--627,
  1995.

\bibitem[Eden and B\"oning(2002)]{eden02-jpo}
C.~Eden and C.~B\"oning.
\newblock Sources of eddy kinetic energy in the {L}abrador {S}ea.
\newblock \emph{J. Phys. Oceanogr.}, 32\penalty0 (12):\penalty0 3346--3363,
  2002.

\bibitem[Gelderloos et~al.(2011)Gelderloos, Katsman, and
  Drijfhout]{gelderloos11-jpo}
R.~Gelderloos, C.~A. Katsman, and S.~S. Drijfhout.
\newblock Assessing the role of three eddy types in restratifying the
  {L}abrador {S}ea after deep convection.
\newblock \emph{J. Phys. Oceanogr.}, 41\penalty0 (11):\penalty0 2102--2119,
  2011.

\bibitem[Kang et~al.(2014)Kang, Belu{\v s}i{\'c}, and Smith-Miles]{kang14-jas}
Y.~Kang, D.~Belu{\v s}i{\'c}, and K.~Smith-Miles.
\newblock Detecting and classifying events in noisy time series.
\newblock \emph{J. Atmos. Sci.}, 71\penalty0 (3):\penalty0 1090--1104, 2014.

\bibitem[Kang et~al.(2015)Kang, Belu{\v s}i{\'c}, and
  Smith-Miles]{kang15-qjrms}
Y.~Kang, D.~Belu{\v s}i{\'c}, and K.~Smith-Miles.
\newblock Classes of structures in the stable atmospheric boundary layer.
\newblock \emph{Q. J. Roy. Meteor. Soc.}, 141\penalty0 (691):\penalty0
  2057--2069, 2015.

\bibitem[{Le H\'{e}naff} et~al.(2014){Le H\'{e}naff}, Kourafalou, Dussurget,
  and Lumpkin]{lehenaff14-pio}
M.~{Le H\'{e}naff}, V.~H. Kourafalou, R.~Dussurget, and R.~Lumpkin.
\newblock Cyclonic activity in the eastern {G}ulf of {M}exico: Characterization
  from along-track altimetry and {\em in situ} drifter trajectories.
\newblock \emph{Prog. Oceanogr.}, 120:\penalty0 120--138, 2014.

\bibitem[Lilly and Gascard(2006)]{lilly06-npg}
J.~M. Lilly and J.-C. Gascard.
\newblock Wavelet ridge diagnosis of time-varying elliptical signals with
  application to an oceanic eddy.
\newblock \emph{Nonlinear Proc. Geoph.}, 13:\penalty0 467--483, 2006.

\bibitem[Lilly and Olhede(2009{\natexlab{a}})]{lilly09-asilomar}
J.~M. Lilly and S.~C. Olhede.
\newblock Wavelet ridge estimation of jointly modulated multivariate
  oscillations.
\newblock In \emph{2009 Conference Record of the Forty-Third Asilomar
  Conference on Signals, Systems, and Computers}, pages 452--456,
  2009{\natexlab{a}}.
\newblock \doi{10.1109/ACSSC.2009.5469858}.
\newblock Refereed conference proceeding paper.

\bibitem[Lilly and Olhede(2009{\natexlab{b}})]{lilly09-itsp}
J.~M. Lilly and S.~C. Olhede.
\newblock Higher-order properties of analytic wavelets.
\newblock \emph{IEEE T. Signal Proces.}, 57\penalty0 (1):\penalty0 146--160,
  2009{\natexlab{b}}.
\newblock \doi{10.1109/TSP.2008.2007607}.

\bibitem[Lilly and Olhede(2010)]{lilly10-itit}
J.~M. Lilly and S.~C. Olhede.
\newblock On the analytic wavelet transform.
\newblock \emph{IEEE T. Inform. Theory}, 56\penalty0 (8):\penalty0 4135--4156,
  2010.

\bibitem[Lilly and Olhede(2012{\natexlab{a}})]{lilly12a-itsp}
J.~M. Lilly and S.~C. Olhede.
\newblock Analysis of modulated multivariate oscillations.
\newblock \emph{IEEE T. Signal Proces.}, 60\penalty0 (2):\penalty0 600--612,
  2012{\natexlab{a}}.

\bibitem[Lilly and Olhede(2012{\natexlab{b}})]{lilly12b-itsp}
J.~M. Lilly and S.~C. Olhede.
\newblock Generalized {M}orse wavelets as a superfamily of analytic wavelets.
\newblock \emph{IEEE T. Signal Proces.}, 60\penalty0 (11):\penalty0 6036--6041,
  2012{\natexlab{b}}.

\bibitem[Lilly and Park(1995)]{lilly95-gji}
J.~M. Lilly and J.~Park.
\newblock Multiwavelet spectral and polarization analysis.
\newblock \emph{Geophys. J. Int.}, 122:\penalty0 1001--1021, 1995.

\bibitem[Lilly et~al.(2003)Lilly, Rhines, Schott, Lavender, Lazier, Send, and
  D'Asaro]{lilly03-pio}
J.~M. Lilly, P.~B. Rhines, F.~Schott, K.~Lavender, J.~Lazier, U.~Send, and
  E.~D'Asaro.
\newblock Observations of the {L}abrador {S}ea eddy field.
\newblock \emph{Prog. Oceanogr.}, 59\penalty0 (1):\penalty0 75--176, 2003.

\bibitem[Lilly et~al.(2017)Lilly, Sykulski, Early, and Olhede]{lilly17-npg}
J.~M. Lilly, A.~M. Sykulski, J.~J. Early, and S.~C. Olhede.
\newblock Fractional {B}rownian motion, the {M}at\'ern process, and stochastic
  modeling of turbulent dispersion.
\newblock 2017.
\newblock URL \url{http://arxiv.org/pdf/1605.01684v1.pdf}.
\newblock Submitted.

\bibitem[Luo et~al.(2011)Luo, Bracco, and Di~Lorenzo]{luo11-pio}
H.~Luo, A.~Bracco, and E.~Di~Lorenzo.
\newblock The interannual variability of the surface eddy kinetic energy in the
  {L}abrador {S}ea.
\newblock \emph{Prog. Oceanogr.}, 91\penalty0 (3):\penalty0 295--311, 2011.

\bibitem[Mallat(1999)]{mallat}
S.~Mallat.
\newblock \emph{A wavelet tour of signal processing, 2nd edition}.
\newblock Academic Press, New York, 1999.

\bibitem[Mallat and Hwang(1992)]{mallat92-itit}
S.~Mallat and W.~L. Hwang.
\newblock Singularity detection and processing with wavelets.
\newblock \emph{IEEE T. Inform. Theory}, 38\penalty0 (2):\penalty0 617--643,
  1992.

\bibitem[Mandelbrot and {Van Ness}(1968)]{mandelbrot68-siam}
B.~B. Mandelbrot and J.~W. {Van Ness}.
\newblock Fractional {B}rownian motions, fractional noises and applications.
\newblock \emph{SIAM Rev.}, 10\penalty0 (4):\penalty0 422--437, 1968.

\bibitem[Mann and Lees(1996)]{mann96-cc}
M.~E. Mann and J.~M. Lees.
\newblock Robust estimation of background noise and signal detection in
  climatic time series.
\newblock \emph{Climatic Change}, 33\penalty0 (3):\penalty0 409--445, 1996.

\bibitem[Olhede and Walden(2002)]{olhede02-itsp}
S.~C. Olhede and A.~T. Walden.
\newblock Generalized {M}orse wavelets.
\newblock \emph{IEEE T. Signal Proces.}, 50\penalty0 (11):\penalty0 2661--2670,
  2002.

\bibitem[Olhede and Walden(2003)]{olhede03a-prsla}
S.~C. Olhede and A.~T. Walden.
\newblock Polarization phase relationships via multiple {M}orse wavelets. {I}.
  {F}undamentals.
\newblock \emph{P. Roy. Soc. Lond. A Mat.}, 459\penalty0 (A):\penalty0
  413--444, 2003.

\bibitem[Thomas and Foken(2004)]{thomas04-tac}
C.~Thomas and T.~Foken.
\newblock Detection of long-term coherent exchange over spruce forest using
  wavelet analysis.
\newblock \emph{Theor. Appl. Climatol.}, 80\penalty0 (2-4):\penalty0 91--104,
  2004.

\bibitem[Thomson(1982)]{thomson82-ieee}
D.~J. Thomson.
\newblock Spectrum estimation and harmonic analysis.
\newblock \emph{Proc. IEEE}, 70\penalty0 (9):\penalty0 1055--1096, 1982.

\bibitem[Zavala-Hidalgo et~al.(2003)Zavala-Hidalgo, Morey, and
  O'Brien]{zavala-hidalgo03-jpo}
J.~Zavala-Hidalgo, S.~L. Morey, and J.~J. O'Brien.
\newblock Cyclonic eddies northeast of the {C}ampeche {B}ank from altimetry
  data.
\newblock \emph{J. Phys. Oceanogr.}, 33\penalty0 (3):\penalty0 623--629, 2003.

\end{thebibliography}

\label{lastpage}

\newpage  \setcounter{page}{1} \setcounter{equation}{0} 
\renewcommand{\theequation}{S\hskip2pt\arabic{equation}}

\section*{Supplementary text for ``Element analysis'' by J. M. Lilly}

\subsection*{Section S1. Moments and cumulants of generalized Morse wavelets}\label{appendix:moments}

The frequency-domain {\em moments} of the generalized Morse wavelets, which are utilized several times in the main text, are defined as [see \cite{lilly09-itsp}, \S~III-A]
\begin{equation}\label{momentexpansion}
 M_{n;\beta,\gamma}\equiv\frac{1}{2\pi}\int_{0}^{\infty}\omega^n\Psi_{\beta,\gamma}(\omega)\rd\omega= \frac{a_{\beta,\gamma}}{2\pi}\int_0^{\infty}\omega^{\beta+n}\re^{-\omega^\gamma}\rd\omega =
 \frac{a_{\beta,\gamma}}{2\pi \gamma}\Gamma\left(\frac{\beta+1+n}{\gamma}\right)
\end{equation}
where the last expression follows from the change of variables $u=\omega^\gamma$ together with the definition of the gamma function $\Gamma(x) \equiv \int_{0}^{\infty} u^{x-1}\re^{-u} \rd u$.  The moments are the terms in the Taylor series 
\begin{equation}
\psi_{\beta,\gamma}(t) = \sum_{n= 0} ^\infty\frac{(\ri t)^n}{n!}\,M_{n;\beta,\gamma}\label{psimoments}
\end{equation}
from which we see that the zeroth-order moment is the value of the wavelet at its temporal center,
\begin{equation}\psi_{\beta,\gamma}(0)=\frac{1}{2\pi}\int_{0}^{\infty}\Psi_{\beta,\gamma}(\omega)\rd\omega=M_{0;\beta,\gamma}= \frac{a_{\beta,\gamma}}{2\pi \gamma}\Gamma\left(\frac{\beta+1}{\gamma}\right).\label{waveletcenter}
\end{equation}
This expression is used in setting the values of the coefficients $|c_n|$ in the element model.  In the synthetic example of figure~\ref{impulses-morsetrain-noisy}, we choose $|c_n \psi_{1,2}(0)|=2$.  From $a_{\beta,\gamma}=2\left(\re \gamma/\beta\right)^{\beta/\gamma}$ we have $a_{1,2}=2\sqrt{2\re\,}=4.66$, and then  $\psi_{1,2}(0)=4.66/(4\pi)=0.37$, leading to $|c_n|=5.39$ as stated in \S~\ref{section:element:examples}.

Closely related to the wavelet moments are the wavelet {\em cumulants}, which are the coefficients in the expansion of the natural logarithm of the wavelet, such that the wavelet is given by
\begin{equation}
 \psi_{\beta,\gamma}(t) = \exp\left\{\sum_{n= 0} ^\infty\frac{(\ri t)^n}{n!}\,K_{n;\beta,\gamma}\right\}.\label{psicumulants}
\end{equation}
Equating the wavelet's moment and cumulant expansions, we find
\begin{multline}
\re^{K_{0;\beta,\gamma}}\left[1+\ri K_{1;\beta,\gamma}t -\frac{1}{2}\left(K_{2;\beta,\gamma}+K_{1;\beta,\gamma}^2\right)t^2 +\ldots\right] \\= M_{0;\beta,\gamma}\left[1+\ri \frac{M_{1;\beta,\gamma}}{M_{0;\beta,\gamma}}t -\frac{1}{2}\frac{M_{2;\beta,\gamma}}{M_{0;\beta,\gamma}}t^2 +\ldots\right]
\end{multline}
from which we have omitted powers of $t$ higher than third order. Equating powers of $t$ then leads to $M_{0;\beta,\gamma}=\re^{K_{0;\beta,\gamma}}$ together with \begin{equation}
K_{1;\beta,\gamma}\equiv\frac{M_{1;\beta,\gamma}}{M_{0;\beta,\gamma}},\quad\quad K_{2;\beta,\gamma} \equiv\frac{M_{2;\beta,\gamma}}{M_{0;\beta,\gamma}}-\frac{M_{1;\beta,\gamma}^2}{M_{0;\beta,\gamma}^2}.
\end{equation}
Equation (\ref{cumulantdefinitions}) in the main text follows from this expression together with (\ref{momentexpansion}) for the moments. 

\subsection*{Section S2. The transform of a Morse function}\label{appendix:wavelet}
In this section, the expression (\ref{simplephi}) for the Morse wavelet transform of another Morse function is derived.  Here we define $\zeta_{\beta,\mu,\gamma}(\tau,s)$ as the transform of a Morse function with $\rho=1$, and find
\begin{multline}
\label{wavetransofpsio}
\zeta_{\beta,\mu,\gamma}(\tau,s) \equiv 
\int_{-\infty}^{\infty} \frac{1}{s}
\psi_{\beta,\gamma}^*\left(\frac{t-\tau}{s}\right)\psi_{\mu,\gamma}\left(t\right)\,\rd t\\
= \frac{a_{\beta,\gamma}\,a_{\mu,\gamma}}{(2\pi)^2}\int_{0}^{\infty}\int_{0}^{\infty}\int_{-\infty}^{\infty}
(s\omega )^\beta \nu^\mu \re^{-(s\omega )^\gamma}\re^{ - \nu^\gamma} \re^{\ri \omega \tau - \ri(\omega-\nu) t}\,\rd t \,\rd \omega \, \rd\nu 
\\ = \frac{a_{\beta,\gamma}\,a_{\mu,\gamma}}{2\pi}\int_{0}^{\infty}\int_{0}^{\infty}
(s\omega )^\beta \nu^\mu \re^{-(s\omega )^\gamma}\re^{ - \nu^\gamma} \re^{\ri \omega \tau} \delta(\omega-\nu) \,\rd \omega \, \rd\nu 
\\= \frac{a_{\beta,\gamma}\,a_{\mu,\gamma}}{2\pi}\int_{0}^{\infty}
(s\omega )^\beta \omega^\mu \re^{-(s\omega )^\gamma}\re^{ - \omega^\gamma} \re^{\ri \omega \tau}\,\rd \omega 
\\=\frac{ a_{\beta,\gamma}\,a_{\mu,\gamma}}{2\pi} \frac{s^{\beta} }{\left(\sqrt[\gamma]{s^\gamma+1}\,\right)^{\beta+\mu}}\, \int_{0}^{\infty}
\left(\sqrt[\gamma]{s^\gamma+1}\, \omega \right)^{\beta+\mu} \re^{-\left(\sqrt[\gamma]{s^\gamma+1}\,\omega \right)^\gamma} \re^{\ri \omega \tau}\,\rd \omega 
\end{multline}
after substituting the wavelet definition (\ref{waveletdef}) to obtain the second line, and where $\delta(\omega)$ is again the Dirac delta function. However the wavelet $\psi_{\beta,\gamma}(t)$ itself can be rescaled to give
\begin{equation} \label{rescaledtransform}
\frac{1}{\sqrt[\gamma]{s^\gamma+1}}\,\psi_{\beta+\mu,\gamma}\left(\frac{\tau}{\sqrt[\gamma]{s^\gamma+1}}\right)=\frac{a_{\beta+\mu}}{2\pi}\int_{0}^{\infty}
\left(\sqrt[\gamma]{s^\gamma+1}\,\omega \right)^{\beta+\mu} \re^{-\left( \sqrt[\gamma]{s^\gamma+1}\,\omega \right)^\gamma } \re^{\ri \omega \tau}\rd \omega 
\end{equation}
as follows from the wavelet definition (\ref{waveletdef}). Combining the previous two expressions gives  (\ref{simplephi}), as claimed.  Then (\ref{morseofmorse}) follows from a change of variables.

\subsection*{Section S3.  The wavelet spectrum of power-law noise}
In this section we derive (\ref{sigmavss}) for the expected value of the magnitude-squared wavelet transform, or wavelet spectrum, of power-law noise.  To do so we will express the wavelet spectrum of noise in terms of a wavelet moment.  The wavelet spectrum of noise (\ref{expectednoisetransform}) is found to be 
\begin{equation}\label{expectedtransform}
\E\left\{\left|\varepsilon_{\beta,\gamma}(\tau,s)  \right|^2\right\} =  \frac{1}{2\pi}\int_{0}^{\infty} \Psi_{\beta,\gamma}^2(s\omega) A^2\omega^{-2\alpha} \rd \omega=A^2 s^{2\alpha-1} f_{\alpha,\beta,\gamma}
\end{equation}
after employing a change of variables, and with $f_{\alpha,\beta,\gamma}$ given by
\begin{multline}\label{fsteps}
f_{\alpha,\beta,\gamma}\equiv \frac{1}{2\pi}\int_{0}^{\infty} \omega^{-2\alpha} \,\Psi_{\beta,\gamma}^2(\omega)\rd \omega=\frac{a_{\beta,\gamma}^2}{2\pi}\int_{0}^{\infty} \omega^{2\beta-2\alpha}\re^{-2\omega^\gamma} \rd \omega\\=\frac{a_{\beta,\gamma}^2}{2^{(2\beta-2\alpha+1)/\gamma}}\frac{1}{2\pi}\int_{0}^{\infty} \omega^{2\beta-2\alpha}\re^{-\omega^\gamma} \rd \omega
= \frac{a_{\beta,\gamma}^2}{2\pi \gamma} \frac{\Gamma\left(\frac{2\beta-2\alpha+1}{\gamma}\right)}{2^{(2\beta-2\alpha+1)/\gamma}} 
\end{multline}
with the second line following from a second change of variables $\omega\mapsto\omega/2^{1/\gamma}$, and then using the definition of the gamma function as in \S~S1 of the supplementary text. Observe that in order for the integral defining $f_{\alpha,\beta,\gamma}$ to evaluate to a finite value, we must have $\beta>\alpha-\frac{1}{2}$, which keeps the argument of the gamma function in (\ref{simplified}) positive.  Intuitively this means that the decay of the wavelet toward zero frequency, controlled by $\beta$, must be strong enough to overcome the singularity in the Fourier spectrum of the noise $S_\epsilon(\omega)$, resulting in an integrable singularity weaker than $\omega^{-1}$.

The moments of the {\em squared} wavelet, referred to as the {\em energy moments} by \cite{lilly09-itsp}, are defined as [see \S~III-A of \cite{lilly09-itsp}]
\begin{equation}\label{energymoments}
 N_{n;\beta,\gamma}\equiv\frac{1}{2\pi}\int_{0}^{\infty}\omega^n\Psi_{\beta,\gamma}^2(\omega)\rd\omega= \frac{a_{\beta,\gamma}^2}{2\pi}\int_0^{\infty}\omega^{2\beta+n}\re^{-2\omega^\gamma}\rd\omega
\end{equation}
so that $f_{\alpha,\beta,\gamma}$ can alternately be expressed in terms of the energy moments as $f_{\alpha,\beta,\gamma}=N_{-2\alpha;\beta,\gamma}$.  This is how $f_{\alpha,\beta,\gamma}$  is implemented in the \texttt{jLab} toolbox.

\subsection*{Section S4. The wavelet transform autocovariance function}
  
In this appendix, we derive (\ref{covresult}) expressing the normalized wavelet transform autocovariance function $\Xi_{\alpha,\beta,\gamma}(\tau,s,r) $ in terms of a wavelet. To show this, we substitute the frequency-domain form of the wavelet transform (\ref{wavetrans}) into the definition (\ref{covdef}), giving 
\begin{equation}
\Xi_{\alpha,\beta,\gamma}(u,s,r)  =  \frac{1}{2\pi} \int_0^\infty \re^{-\ri\omega u}
\Psi_{\beta,\gamma}(s\omega) \Psi_{\beta,\gamma}(rs\omega) S_\epsilon(\omega) \rd\omega
\end{equation}
after employing the definition of the noise spectrum (\ref{noisespectrumdefinition}). Inserting the form of the power-law noise spectrum $S_\epsilon(\omega) = A^2 \omega^{-2\alpha}$, this expression becomes 
\begin{equation}
\Xi_{\alpha,\beta,\gamma}(u,s,r)  =A^2 a_{\beta,\gamma}^2
\frac{1}{2\pi} \int_0^\infty (s\omega)^\beta (rs\omega)^\beta \omega^{-2\alpha} \re^{-(s\omega)^\gamma-(rs\omega)^\gamma}\re^{-\ri\omega u} \rd\omega 
\end{equation}
using the wavelet definition (\ref{waveletdef}). In terms of $\tilde r_\gamma\equiv\sqrt[\gamma]{1+r^\gamma}$ this may be rewritten as
\begin{equation}\label{intermediatexi}
\Xi_{\alpha,\beta,\gamma}(u,s,r)  =\frac{ A^2 a_{\beta,\gamma}^2}{a_{2\beta-2\alpha,\gamma}} \frac{r^{\beta} s^{2\alpha-1}}{\tilde r_\gamma^{2\beta-2\alpha+1}} \left[
\frac{s\tilde r_\gamma}{2\pi} \int_0^\infty a_{2\beta-2\alpha,\gamma} (s\tilde  r_\gamma \omega)^{2\beta-2\alpha} \re^{-(s\tilde r_\gamma \omega)^\gamma}\re^{-\ri\omega u} \rd\omega \right]
\end{equation}
however, from the wavelet scaling law
\begin{equation}
\psi_{\beta,\gamma}(t/s) =\frac{s}{2\pi } \int_{-\infty}^{\infty} \Psi_{\beta,\gamma}(s\omega) \re^{\ri \omega t}\,\rd \omega = \frac{s}{2\pi } \int_{-\infty}^{\infty}  a_{\beta,\gamma}(s\omega)^\beta \re^{-(s\omega)^\gamma}\re^{\ri \omega t}\,\rd \omega
\end{equation}
one sees that the quantity in brackets in (\ref{intermediatexi}) is the same as $\psi_{2\beta-2\alpha,\gamma}^*\left(u/(s\tilde r_\gamma)\right)$, leading to 
\begin{equation}
\Xi_{\alpha,\beta,\gamma}(u,s,r)  = A^2 \frac{a_{\beta,\gamma}^2}{a_{2\beta-2\alpha,\gamma}} \frac{r^{\beta} s^{2\alpha-1}}{\tilde r_\gamma^{2\beta-2\alpha+1}}\,\psi_{2\beta-2\alpha,\gamma}^*\!\left(\frac{u}{s\tilde r_\gamma }\right).
\end{equation}
This can be simplified further by making use of the final expression for $f_{\alpha,\beta,\gamma}$ in  (\ref{fsteps}), together with (\ref{sigmavss}) for  $\sigma_{\alpha,\beta,\gamma}^2(s)$.  These two substitutions lead to (\ref{covresult}), as claimed.

Recalling from (\ref{waveletcenter}) that the value of the $(2\beta-2\alpha,\gamma)$ wavelet at its temporal center is given by
\begin{equation}\label{simplifiedfdef}
\psi_{2\beta-2\alpha,\gamma}(0)=\frac{a_{2\beta-2\alpha,\gamma}}{2\pi \gamma}\Gamma\left(\frac{2\beta-2\alpha+1}{\gamma}\right)
\end{equation}
one finds the following simplification for the ratio of this wavelet to the $f_{\alpha,\beta,\gamma}$ function
\begin{equation}\label{simplifiedfdef}
\frac{\psi_{2\beta-2\alpha,\gamma}(0)}{f_{\alpha,\beta,\gamma}}=\frac{a_{2\beta-2\alpha,\gamma}}{a^2_{\beta,\gamma}} 2^{(2\beta-2\alpha+1)/\gamma}. 
\end{equation}
Substituting this expression into (\ref{covresult}), we find $\Xi_{\alpha,\beta,\gamma}(0,s,1) =\sigma^2_{\alpha,\beta,\gamma}(s)$, as claimed.

\newpage  \setcounter{page}{1} \setcounter{figure}{0} 
\renewcommand{\thefigure}{S\arabic{figure}}

\section*{Supplementary figures for ``Element analysis'' by J. M. Lilly}

\begin{figure}[h!]
\includegraphics[width=\textwidth]{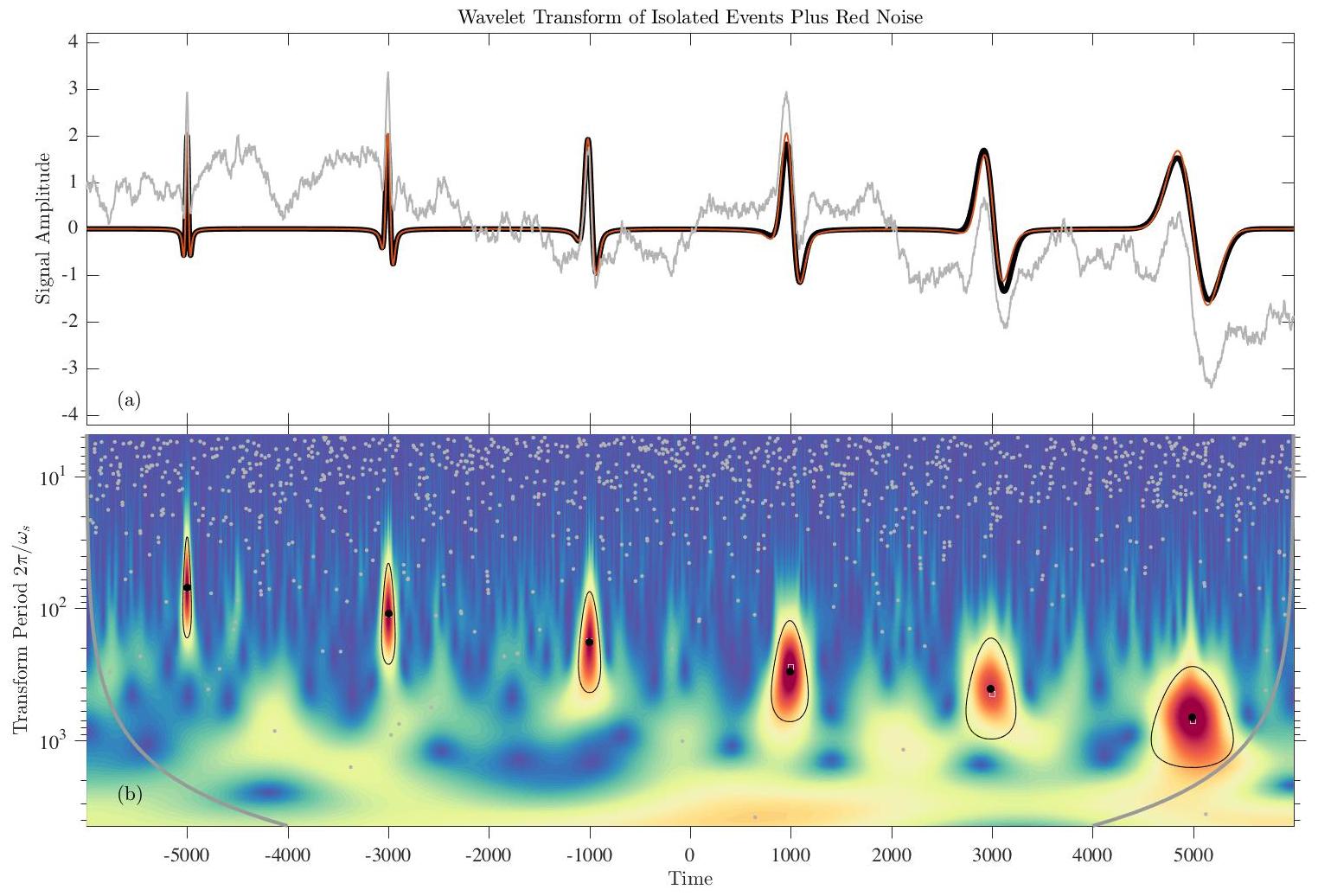}
\caption{As with figure~\ref{impulses-morsetrain-noisy} in the main text, but with the noise consisting of a realization of $\alpha=1$ red noise rather than white noise.  The red noise is formed by cumulatively summing discrete Gaussian white noise, then setting the standard deviation to unity and removing the mean.   All other aspects of this plot are identical with figure~\ref{impulses-morsetrain-noisy}.  }\label{impulses-morsetrain-rednoisy}
\end{figure}

\begin{figure}[h!]
\includegraphics[width=\textwidth]{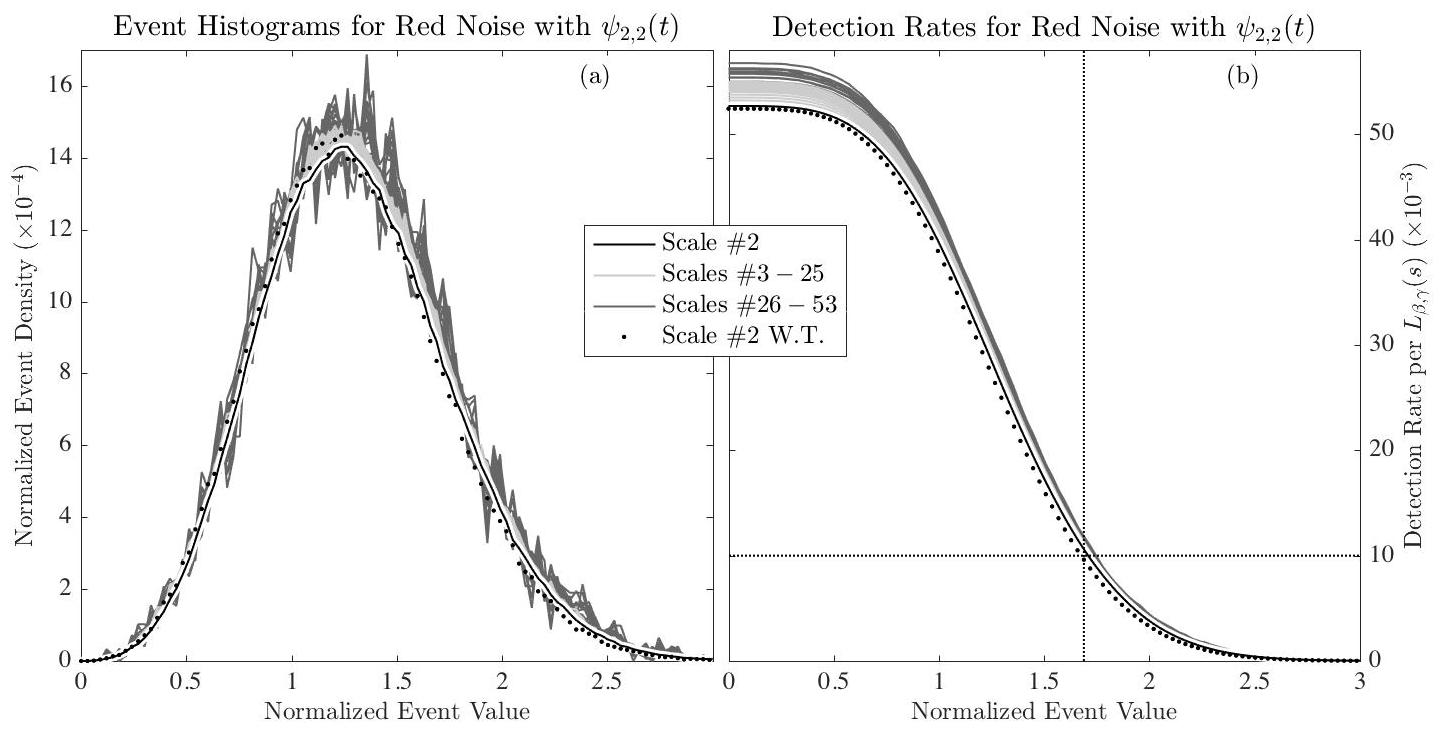}
\caption{As with figure~\ref{impulses-noisedist-22} in the main text, but for  $\alpha=1$ red noise.  The differences between the normalized curves in the white and red noise cases are marginal, and are primarily limited to the magnitudes of event densities and detection rates rather than the shapes of the curves; note that the $y$-axes here are different from those in figure~\ref{impulses-noisedist-22}.}
\end{figure}

\begin{figure}[h!]
\includegraphics[width=1\textwidth]{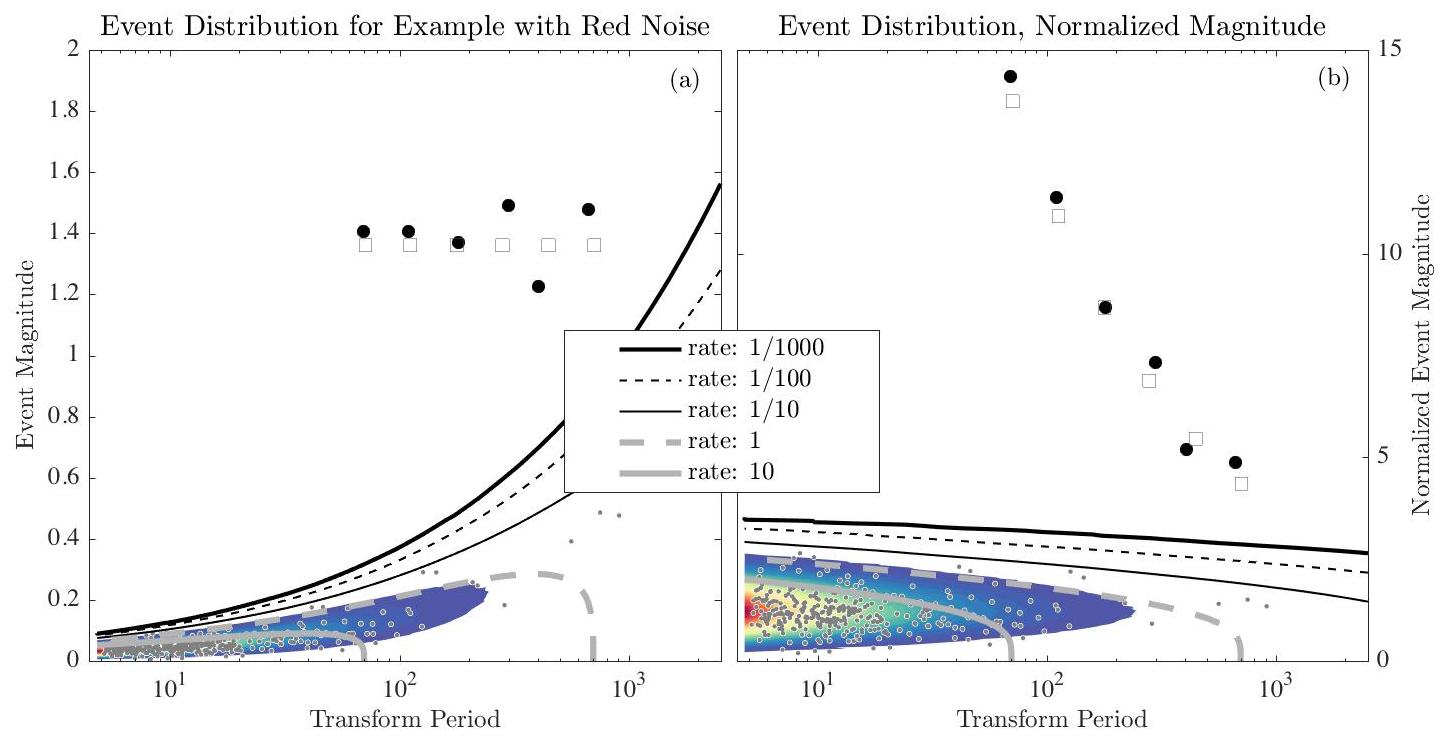}
\caption{As with figure~\ref{impulses-examplenoisedist} in the main text, but for  $\alpha=1$ red noise.  Unlike in the white noise case, detected event magnitudes due entirely to the noise tend to increase with increasing scale or period, as is clear from panel (a).  Normalization by the wavelet spectrum of the noise, as in (b), makes the noise distributions qualitatively similar between the red and white noise cases; however, there is now the important difference that events of a fixed magnitude, such as the six large-amplitude events in the analyzed signal, becomes {\em less} rather than more significant as one proceeds toward larger scales or periods.  This figure is used to assess the significance of events in figure~\ref{impulses-morsetrain-rednoisy}.  As with the white noise case, the six large-amplitude events are again highly statistically significant.}
\end{figure}

\begin{figure}[h!]
\includegraphics[width=\textwidth]{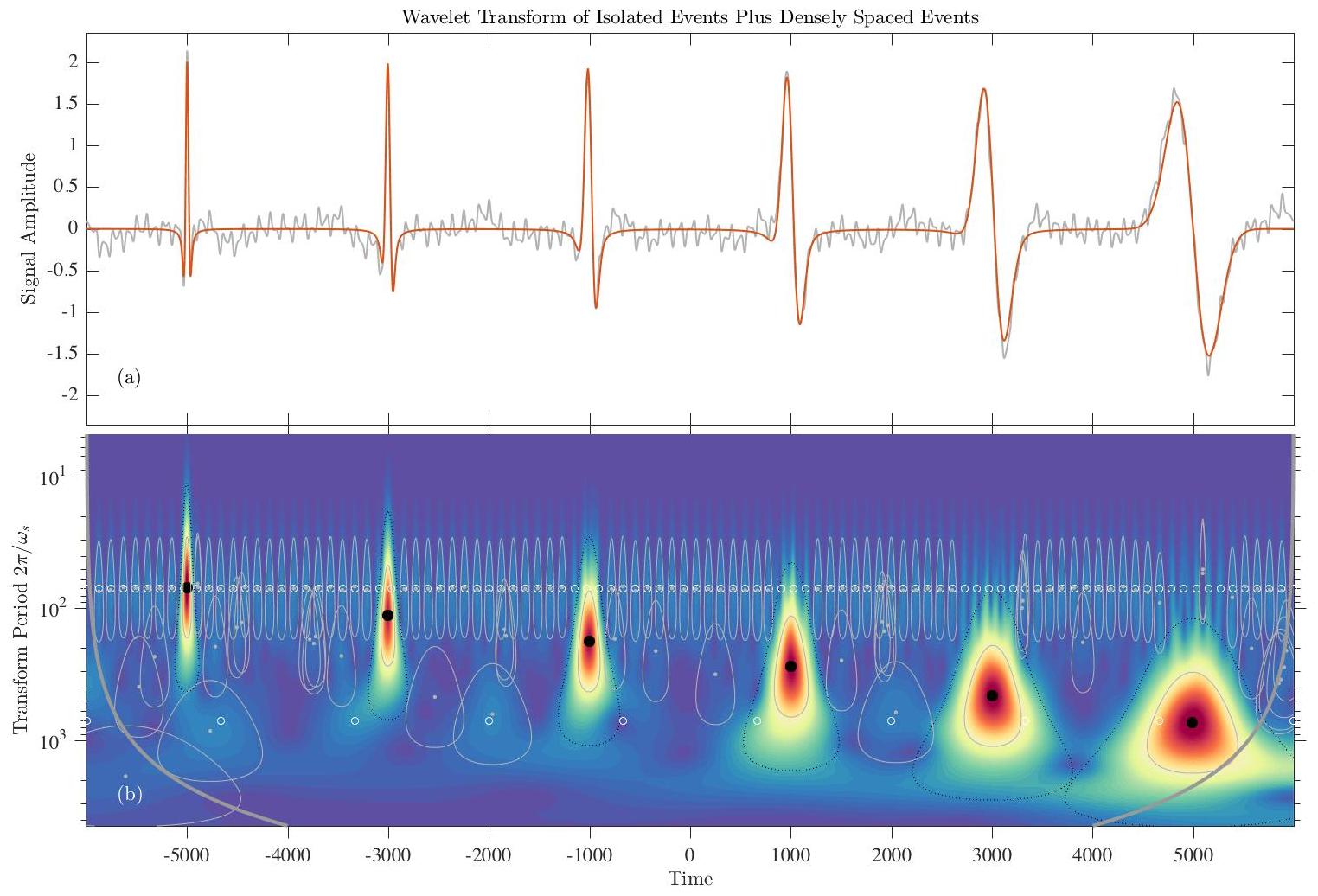}
\caption{As with figure~\ref{impulses-morsetrain-noisy}, but for the chain of six events plus an array of closely spaced, weaker events rather than white noise.   This figure explores the sensitivity of event detection to the presence of nearby, stronger events.  The red line in (a) shows the signal due to the six major events.  To this is added a set of 110 smaller-amplitude events, with the total signal shown as a gray line. (Unlike figure~\ref{impulses-morsetrain-noisy}, no reconstruction is shown in (a), because the focus here is on detection.)  All signals are the real parts of phase-rotated $\psi_{1,2}(t)$ wavelets.  The peak magnitudes of all of smaller-amplitude events is set to 1/10 that of the six major events, while their phases are random.  100 of these have a frequency of $\omega_\rho=2\pi/100$, while then remaining ten have $\omega_\rho=2\pi/1000$; both sets are uniformly spaced in time.  Using (\ref{rhotos}), these peak frequencies are found correspond to scale frequencies within the transform of $\omega_s=\sqrt{2}\omega_\rho$, or  $\omega_s=2\pi\sqrt{2}/100$ and  $\omega_s=2\pi\sqrt{2}/1000$ respectively.  The locations of the small-amplitude events are marked in (b) with the small white circles, and occur in two parallel rows.  In  (b), the detected large-amplitude events are again marked as solid black circles, while the detected small-amplitude events are marked as gray dots.  Gray curves show the $\lambda=1/2$ regions of influence, see \S~\ref{subsection:region}, around each detected maximum, while black dotted lines show the larger  $\lambda=1/10$ regions around just the large-amplitude maxima.  The purpose of this figure is to compare the true locations of the smaller-amplitude maxima, marked by the white circles, with the inferred locations as marked by the gray dots surrounded by gray curves. When the smaller-amplitude maxima are sufficiently distant from the large-amplitude maxima, they are accurately detected.  However, those in the vicinity of the larger-amplitude maxima are obscured.  As the smaller-amplitude maxima are 1/10 as strong as the larger-amplitude maxima, the obscured region extends past the location at which the larger-amplitude maxima have decayed to 1/10 of their value, marked by the dotted black curves.  Spurious maxima, mostly arising from the interaction between the larger and smaller-scale layers of the smaller-amplitude maxima, are also seen.  Some of these would be removed by the isolation criterion, which has not been applied in this plot.  This illustrates some limitations of element analysis due to the interaction of nearby elements. }
\end{figure}

\end{document}